\documentclass[%
aps,
pre,
 reprint,
 showkeys,
 superscriptaddress,
 amsmath,amssymb,
 aps,
floatfix,
]{revtex4-2}

\usepackage{graphicx}
\usepackage{dcolumn}
\usepackage{bm}
\usepackage{hyperref}
\usepackage{siunitx}
\begin{document}

\preprint{APS/123-QED}

\title{Feedback between microscopic activity and macroscopic dynamics drives excitability and oscillations in mechanochemical matter}
\author{Tim Dullweber}
    \affiliation{%
European Molecular Biology Laboratory, Meyerhofstraße 1, 69117 Heidelberg\\
}%
    \affiliation{%
Department of Physics and Astronomy, Heidelberg University, 69120 Heidelberg
}%
\author{Roman Belousov}
     \affiliation{%
European Molecular Biology Laboratory, Meyerhofstraße 1, 69117 Heidelberg\\
}%
\author{Anna Erzberger}%
 \email{erzberge@embl.de}
    \affiliation{%
European Molecular Biology Laboratory, Meyerhofstraße 1, 69117 Heidelberg\\
}%
    \affiliation{%
Department of Physics and Astronomy, Heidelberg University, 69120 Heidelberg
}%

\date{\today}

\begin{abstract}
The macroscopic behaviour of active matter arises from nonequilibrium microscopic processes. In soft materials, active stresses typically drive macroscopic shape changes, which in turn alter the geometry constraining the microscopic dynamics, leading to complex feedback effects. Although such mechanochemical coupling is common in living matter and associated with biological functions such as cell migration, division, and differentiation, the underlying principles are not  well understood due to a lack of minimal models that bridge the scales from the microscopic biochemical processes to the macroscopic shape dynamics. 
To address this gap, we derive tractable coarse-grained equations from microscopic dynamics for a class of mechanochemical systems, in which biochemical signal processing is coupled to shape dynamics. Specifically, we consider molecular interactions at the surface of biological cells that commonly drive cell-cell signaling and adhesion, and obtain a macroscopic description of cells as signal-processing droplets that adaptively change their interfacial tensions.
We find a rich phenomenology, including multistability, symmetry-breaking, excitability, and self-sustained shape oscillations, with the underlying critical points revealing universal characteristics of such systems. 
Our tractable framework provides a paradigm for how soft active materials respond to shape-dependent signals, and suggests novel modes of self-organisation at the collective scale. These are explored further in our companion manuscript ~\cite{PRLJoint}.

\end{abstract}

\maketitle
\section{Introduction}
Active materials exhibit behaviors such as stress generation, sustained oscillations, nonequilibrium phase transitions, and spontaneous symmetry-breaking~
\cite{fodor2018statistical, needleman2017active, ramaswamy2017active, marchetti2013hydrodynamics}. These phenomena result from changes of macroscopic system states driven by microscopic chemical activity \cite{ray2023rectified, goldbeter2018dissipative}. Living materials, in particular, remain far from equilibrium through the continuous production, degradation, and active transport of molecules, leading to spatially varying concentration fields and dynamically evolving material properties \cite{singh2024non}.

By controlling the concentration and spatial distribution of adhesion molecules, molecular motors, cytoskeletal components, and their regulators, cells adapt their mechanical properties---such as elasticity, viscosity, and wettability---in response to changing environmental conditions \cite{maitre2013three, papusheva2010spatial, kasza2007cell}.

\emph{Mechanochemical feedback} arises when these coarse-grained material properties in turn control the chemical composition or spatial distribution of molecular constituents~\cite{rombouts2023forceful,chan2017coordination,gross2017active}. For instance, mechanical stresses can affect the synthesis of new molecules in cells~\cite{dupont2022mechanical}; active hydrodynamic flows control the transport of cytoskeletal components and molecular motors~\cite{shamipour2021cytoplasm, gross2017active, nishikawa2017controlling, mayer2010anisotropies}; active stresses lead to the disassembly of macro-molecular complexes \cite{vogel2013myosin}. Moreover deformations and shape changes can directly impact the microscopic dynamics by changing the domain on which these processes evolve~\cite{wurthner2023geometry, burkart2022control, datta2022assemble, tamemoto2020pattern, mietke2019self, mietke2019minimal}.

Environmental signals are often detected through biochemical reactions at the cell surface, for example via binding of external ligand molecules to receptors, which change the bulk concentration of proteins by regulating their production through gene transcription and translation~\cite[Chap.~15]{alberts2022molecular}. Such signaling interactions depend on the geometry and duration of cell-cell contacts in various contexts~\cite{dullweber2023mechanochemical, barone2017effective, shaya2017cell, khait2016quantitative}. For example, when signaling molecules bind to receptors at cell-cell or cell-substrate interfaces, the resulting response can depend on the available contact area. Indeed, the strength of Delta-Notch signaling interactions \cite{sprinzak2021biophysics}, in which membrane-bound Delta ligands bind to Notch receptors, triggering the transcriptional regulation of various target genes, has been shown to correlate with the area of cell-cell contacts in different experiments \cite{guisoni2017diversity, shaya2017cell,khait2016quantitative}. 
Moreover, Notch signals often trigger changes in mechanical regulators, including cell-cell adhesion molecules, leading to mechanochemical feedback loops \cite{dullweber2023mechanochemical}, which coordinate fate patterning in embryonic development, regeneration, and homeostasis \cite{cohen2023precise, cohen2022mechano,  dray2021dynamic, erzberger2020mechanochemical, priya2020tension}, and which have been successfully engineered synthetically \cite{toda2018programming}.

Identifying the key physical principles of mechanochemical self-organization presents a challenge, because the relevant processes are coupled across different temporal and spatial scales. To bridge the gap between the \emph{microscopic} molecular interactions and the \emph{macroscopic} shape and state dynamics in signal-processing active matter, we take advantage of the time-scale separation that arises generically when biochemical signals trigger adaptive responses through transcriptional regulation \cite[Chap.~15]{alberts2022molecular}. These responses are controlled by molecular production and decay processes which evolve on a slow timescale (from \SI{10}{\min} to \SI{1}{\hour} or longer~\cite{SHAMIR20161302, milo2015cell}) compared to the typical speed of cell-shape changes ($<$\SI{1}{\min}~\cite{wyatt2016question, tran1991time}). Specifically, from a detailed biophysical model of Delta-Notch signaling and adaptive cell-cell adhesion, we derive a minimal set of macroscopic equations, which represent cells as signal-processing \emph{droplets} with adaptive interfacial tensions.
We find that the coupling between active mechanics and biochemical signaling gives rise to dynamical phenomena such as bistability, excitability, and tunable oscillations of droplet shapes and internal states. By identifying the underlying codimension-1 and -2 bifurcations, we reveal universal macroscopic characteristics of such mechanochemical systems.
Our companion manuscript extends the macroscopic framework to collective droplet configurations and a broader range of signaling interactions, and contains an application to experimental data from zebrafish embyros~\cite{PRLJoint}.


\section{Microscopic dynamics of adhesion and signaling}
\label{sec:microscopic_dynamics}

In the following, we derive equations for the dynamics of molecules that can mediate (i) adhesion at contact surfaces, and (ii) the exchange of chemical signals, in the  bulk $\Omega$ and at the surface $\Gamma$ of biological cells [Fig.~\ref{fig:1}], starting from general arguments.
\subsection{Reaction-diffusion equations with surface-bulk coupling}
\label{sec:microscopic_dynamics_general}

\begin{figure*}
    \centering
    \includegraphics[width=\textwidth]{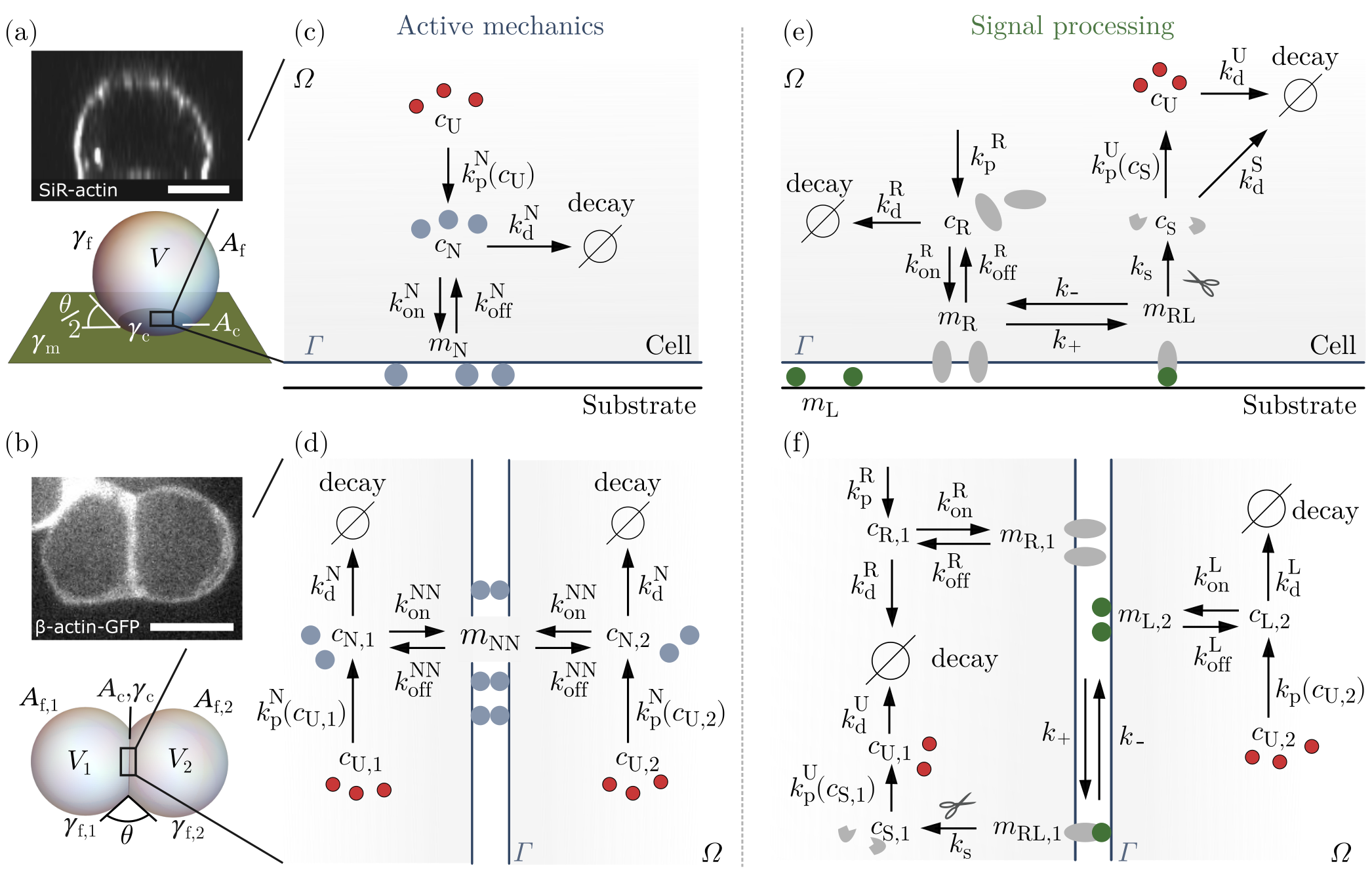}
    \caption{Microscopic interactions underlying adhesion and signaling. (a-b) Schematics of (a) a cell adhering to a substrate, and (b) a cell pair with a shared interface, with fixed volumes $V,V_{1,2}$ and uniform interfacial tensions $\gamma_{\rm{c}}, \gamma_{\rm{m}}, \gamma_{\rm{f,1,2}}$ conjugate to interfacial areas $A_{\rm{c}}, A_{\rm{m}}, A_{\rm{f,1,2}}$. Fluorescence microscopy images show (a) a 3T3 fibroblast on a micropatterned substrate (Appendix~\ref{sec:experimental_methods}) and (b) a pair of zebrafish sensory cells exchanging Notch signals across their contact surface (adapted from \cite{erzberger2020mechanochemical}). Scale bars: \SI{5}{\micro\meter}. (c-d) Adhesion molecules (N, blue) are exchanged between bulk and surface and form complexes at the cell-substrate (c) or across the cell-cell interface (d), which reduce the surface energy. The rate of adhesion molecule production due to transcriptional regulation depends on the regulator concentration $c_\mathrm{U}$ (red); decay rates are constant. (e-f) Contact-dependent signals are received from a ligand-coated substrate (e) or exchanged at the cell-cell contact (f). Receptor (R, gray) and ligand (L, green) molecules bind across the interface and form receptor-ligand complexes (RL). In the single cell (e), receptors are produced in the bulk, exchanged with the surface and bind to substrate-bound ligands, while in cell pairs (f), both receptors and ligands bind and unbind from the cell-cell interface. Receptor-ligand complexes are cleaved irreversibly \cite{bray2016notch}, which releases a signaling molecule (S, gray) into the bulk. Regulator molecules (U, red) are produced with a rate depending on the bulk concentration of signal molecules, and in turn determine the production rate of new ligands. Substrate-bound ligand molecules are released upon the cleavage event and can bind a new receptor, while in  cell pairs, the remaining part of the RL complex is degraded after cleavage \cite{bray2016notch}. Here $c$ and $m$ denote bulk and surface concentrations respectively, and $k$ denote the kinetic rates of the reactions.}
    \label{fig:1}
\end{figure*}
 
We consider continuity equations for the particle densities of these molecules within the bulk $c$ and at the surface $m$ of the form \cite{Hausberg:2018aa}
\begin{align}
    \partial_t c &= D_{\mathrm{c}} \nabla^2 c + \mathcal{R}_{\mathrm{c}} \label{eq:continuity_bulk}\\
    \partial_t m &= D_{\mathrm{m}} \nabla^2 m + \mathcal{R}_{\mathrm{m}} \label{eq:continuity_surface}
\end{align}
in which we consider a diffusive flux with coefficients $D_{\rm{c}}, D_{\rm{m}}$ in three and two dimensions respectively, and reaction terms $\mathcal{R}_{\rm{c}}$ and $\mathcal{R}_{\rm{m}}$. We do not consider convective flows or other active transport processes here. Bulk and surface densities are coupled by the boundary condition
\begin{equation}
    -D_{\mathrm{c}} (\mathbf{n}\cdot\nabla) \left.c\right|_\Gamma =j
    \label{eq:bc_flux}
\end{equation}
in which $j$ is the flux between bulk and surface and $\mathbf{n}$ is the normal vector to the surface pointing outwards. A simple form of this flux is given by \cite{brauns2020phase, khait2016quantitative} 
\begin{equation}
   j = \left. k_{\mathrm{on}} c \right|_{\Gamma} - k_{\mathrm{off}} m
    \label{eq:flux}
\end{equation}
with $k_{\mathrm{on}}$ setting the rate with which molecules bind to the surface and $k_{\mathrm{off}}$ the rate with which they are released from the surface into the bulk.
Reactions in the bulk follow \cite{ross2021proteome}
\begin{align}
    \mathcal{R}_{\mathrm{c}} = k_{\rm{p}} - k_{\rm{d}} c
    \label{eq:bulk_reaction}
\end{align}
with $k_{\mathrm{p}}$ describing an active production of molecules (e.g. due to protein translation in cells) and $k_{\mathrm{d}}$ their rate of decay (protein degradation). The bulk production of molecules drives the system out of thermodynamic equilibrium. For the surface reactions $\mathcal{R}_{\mathrm{m}}$, we consider different molecular processes governing adhesion and contact-based signaling, as specified in the following sections.

Averaging Eq.~\eqref{eq:continuity_bulk} over the bulk's volume $V$ and using Eq.~\eqref{eq:bc_flux} and \eqref{eq:bulk_reaction} yields the dynamic equation for the average bulk concentration $\langle c \rangle$
\begin{align}
    \frac{d \langle c \rangle}{dt} = 
    k_{\mathrm{p}} 
    - k_{\mathrm{d}} \langle c \rangle 
    - \frac{1}{V} \int_{\Gamma} j dA.
    \label{eq:bulk_average_dynamic}
\end{align}
We define the steady state average density in the absence of boundary flux as the reference density $c^0=k_{\mathrm{p}}/k_{\mathrm{d}}$, and defining $m^0=k_{\mathrm{on}}c^0/k_{\mathrm{off}}$ permits introducing normalized particle densities
$c/c^{\mathrm{0}}$ and $m/m^{\mathrm{0}}$. 

With diffusion timescale $\tau_{\mathrm{D}}=V^{2/3}/D_{\mathrm{c}}$ and reaction timescales $\tau_{\mathrm{R}}=1/k_{\mathrm{d}}$ and $\tau_{\mathrm{on}}=V^{1/3}/k_{\mathrm{on}}$, Eq.~\eqref{eq:continuity_bulk} and Eq.~\eqref{eq:bc_flux} with time rescaled in units of $\tau = t/\tau_{\rm{R}}$ read
\begin{align}
    \frac{\tau_{\mathrm{D}}}{\tau_{\rm{R}}} \frac{ \partial_{\tau} c}{c^0} & = V^{2/3} \nabla^2 \frac{c}{c^0} + \frac{\tau_{\mathrm{D}}}{\tau_\mathrm{R}} \left(1 - \frac{c}{c^0} \right), \label{eq:continuity_bulk_timescales}\\
       \left(\mathbf{n}\cdot\nabla\right) \left. \frac{c}{c^0}\right|_\Gamma & = \frac{\tau_{\mathrm{D}}}{\tau_{\mathrm{on}} V^{1/3}}\left( \frac{m}{m^0}-\frac{c}{c^0} \right).
       \label{eq:bc_flux_timescales}
\end{align}
Inside a cell, a typical diffusion constant for a monomeric protein is about \SI{10}{\micro\metre\squared\per\second}, i.e. it takes roughly 10 seconds for a protein to traverse a eukaryotic cell \cite{milo2015cell}. 
Most biochemical reactions are catalyzed by enzymes and occur within less than a second \cite{SHAMIR20161302}, however transcription and translation, i.e. the synthesis of new proteins, and protein degradation take minutes to hours and can vary greatly between different proteins \cite{buccitelli2020mrnas, SHAMIR20161302}. This work focuses on the long time scale dynamics dominated by such production and decay processes, thus, we consider the limit in which bulk diffusion is much faster than the reaction kinetics, i.e.  $\tau_{\mathrm{D}} \ll \tau_{\mathrm{R}}$ and $\tau_{\mathrm{D}} \ll \tau_{\mathrm{on}}$. In this limit, the boundary condition Eq.~\eqref{eq:bc_flux_timescales} is reflective and Eq.~\eqref{eq:continuity_bulk_timescales} becomes a Laplace equation that is solved by a uniform concentration set by the solution of Eq.~\eqref{eq:bulk_average_dynamic} (\emph{shadow limit} \cite{hausberg2018well}).  

The surface of a cell, which is in contact with another cell or a substrate, can be separated into the domain of the contact interface $\Gamma_{\rm{c}}$ and the free surface $\Gamma_{\rm{f}}$ [Fig.~\ref{fig:1}(a)]. At the free surface, molecules can be exchanged with the bulk, but interactions between adhesion or signaling molecules are restricted to the contact interface. The reaction term at the free surface is therefore
$\left.\mathcal{R}_{m}\right|_{\Gamma_\mathrm{f}} =j$.

We assume in the following that the contact line separating the two surfaces forms a diffusive barrier, i.e. that no molecules can diffuse laterally between the surfaces. This assumption substantially simplifies our calculations, and indeed diffusion barriers based on protein structures associated with the membrane, the lipid composition or extreme curvatures---as given at the contact line---have been found to impede diffusive transport on cellular membranes \cite{jacobson2019lateral, trimble2015barriers}. The boundary conditions for the surface densities on the two domains then are
\begin{align}
    \left.(\mathbf{n}\cdot\nabla)m\right|_{\partial \Gamma_{\mathrm{f}}} = 0 
    \label{eq:bc_Af}
    \\
    \left.(\mathbf{n}\cdot\nabla)m\right|_{\partial \Gamma_{\mathrm{c}}} = 0
    \label{eq:bc_Ac}
\end{align}
in which $\partial \Gamma_\mathrm{f}, \partial \Gamma_\mathrm{c}$ denote the contact line. From Eqs.~\eqref{eq:continuity_surface} follows at steady state $\left.j\right|_{\Gamma_\mathrm{f}}=0$ and the uniform steady state bulk concentration 
\begin{equation}
    c = 
    \frac{k_{\mathrm{p}}}{k_\mathrm{d}} - \dfrac{1}{k_{\mathrm{d}} V}  {\displaystyle \int_{\Gamma_\mathrm{c}}} j dA
    \label{eq:bulk_concentration_average_Ac}
\end{equation}
only depends on processes in the bulk and at the contact site $\Gamma_{\mathrm{c}}$. 

In the following sections, we introduce the reaction terms and corresponding boundary fluxes for the adhesion and signaling dynamics at the contact site and compute the steady state bulk and surface densities that fulfill Eqs.~\eqref{eq:continuity_surface}, \eqref{eq:bulk_concentration_average_Ac}, and boundary condition $\eqref{eq:bc_Ac}$. Lateral diffusion coefficients of proteins on lipid membranes are variable and on the order of \SIrange{0.01}{10}{\micro\metre\squared\per\second}~\cite{khait2016quantitative,ramadurai2009lateral, sankaran2009diffusion, tveriakhina2024temporal}, allowing density fields to acquire their steady state within seconds over micrometer length-scales, while equilibration takes substantially longer in larger systems, such as synthetic biomimetic droplets \cite{gonzales2023bidirectional}.

\subsection{Modulation of interfacial tensions by adhesion}
\label{sec:adhesion}
[Chap.~19] Similar problems are discussed in~\cite{sackmann2014physics, schwarz2013physics, smith2007vesicles, smith2005effective,seifert1990adhesion}.

\subsubsection{Tension at the contact interface of a droplet wetting a solid substrate}
We first derive the adaptive adhesion term for a contact surface between a fluid droplet and a solid substrate, to which adhesion molecules can bind [Fig.~\ref{fig:1}(a)]. A mass-action based reaction term for the surface concentration  Eq.~\eqref{eq:continuity_surface} of adhesion molecules reads
\begin{equation}
    \mathcal{R}_{m_{\mathrm{N}}} = k_{\mathrm{on}}^{\mathrm{N}}(m_{\mathrm{N}}^{\mathrm{max}} - m_{\mathrm{N}}) c_{\mathrm{N}} - k_{\mathrm{off}}^{\mathrm{N}} m_{\mathrm{N}}
    \label{eq:R_adhesion_singlet}
\end{equation}
with $m_{\mathrm{N}}^{\mathrm{max}}$ the density of available binding sites at the contact. The flux coupling bulk and contact surface is $j_\mathrm{N} = \mathcal{R}_{m_\mathrm{N}}$ and adhesion molecules bound to the substrate are fixed in place, i.e. $D_{m_\mathrm{N}}=0$ in Eq.~\eqref{eq:continuity_surface}. At steady state, it follows from Eqs.~\eqref{eq:continuity_surface}, \eqref{eq:bulk_concentration_average_Ac},\eqref{eq:R_adhesion_singlet} and boundary condition Eq.~\eqref{eq:bc_Ac} that $j_\mathrm{N}=0$, $c_\mathrm{N} = k_\mathrm{p}^\mathrm{N}/k_\mathrm{d}^\mathrm{N}$, and
\begin{align}
    m_{\mathrm{N}} = \dfrac{
    k_{\mathrm{on}}^{\mathrm{N}} k_{\rm{p}}^{\mathrm{N}}
    }{
   k_{\mathrm{on}}^{\mathrm{N}} k_{\rm{p}}^{\mathrm{N}} +
        k_{\mathrm{off}}^{\mathrm{N}} k_{\rm{d}}^{\rm{N}}
        }m_{\mathrm{N}}^{\mathrm{max}}.
    \label{eq:mN}
\end{align}
The same expression can also be derived from the grand canonical ensemble (Appendix~\ref{sec:adhesion_statmech}). Expansion in the dilute limit  $k_{\mathrm{on}}^{\mathrm{N}} k_{\rm{p}}^{\rm{N}}/k_{\rm{d}}^{\rm{N}} \ll k_{\mathrm{off}}^{\mathrm{N}}$, i.e. where saturation effects do not play a role, yields
\begin{align}
   m_{\mathrm{N}} = 
    \frac{
    k_{\mathrm{on}}^{\mathrm{N}} k_{\rm{p}}^\mathrm{N}
    }{
    k_{\mathrm{off}}^{\mathrm{N}} k_{\rm{d}}^{\rm{N}}
    } m_\mathrm{N}^{\mathrm{max}}
    + \mathcal{O}\left(
    \left(
    \frac{
    k_{\mathrm{on}}^{\mathrm{N}} k_{\rm{p}}^{\rm{N}}
    }{
    k_{\mathrm{off}}^{\mathrm{N}} k_{\rm{d}}^{\rm{N}}
    }
    \right)^2
    \right).
    \label{eq:mN_expansion}
\end{align}

Given that each adhesion complex reduces the surface energy by $\epsilon$ \cite{maitre2011role}, the surface tension at a contact site [Fig.~\ref{fig:1}(a)] in this limit is
\begin{equation}
    \gamma_{\mathrm{c}} = \gamma_0 - \epsilon \frac{
     k_{\mathrm{on}}^{\mathrm{N}} k_{\rm{p}}^{\rm{N}}
    }{
    k_{\mathrm{off}}^{\mathrm{N}} k_{\rm{d}}^{\rm{N}}
    } m_\mathrm{N}^{\mathrm{max}}
    \label{eq:contact_tension_singlet},
\end{equation}
with $\gamma_0$ a baseline tension that contains all other components of the interfacial tension.

\subsubsection{Tension at the interface between two droplets}
At contact surfaces between two droplets, adhesion molecules produced within the droplets can bind across the interface and form adhesion complexes with surface density $m_{\mathrm{NN}}$ [Fig.~\ref{fig:1}(d)] \cite{maitre2012adhesion}. Taking exclusion effects into account, adhesion complexes can only form at unoccupied sites on the interface. The density of unoccupied sites is $(m_\mathrm{NN}^\mathrm{max} - m_\mathrm{NN})$ with $m_\mathrm{NN}^\mathrm{max}$ the maximum possible density of adhesion complexes. The reaction term for the density of adhesion complexes is then 
\begin{align}
    \mathcal{R}_{m_{\mathrm{NN}}} = k_{\mathrm{on}}^{\mathrm{NN}}(m_{\mathrm{NN}}^{\mathrm{max}} - m_{\mathrm{NN}})  c_{\mathrm{N},1} c_{\mathrm{N},2} - k_{\mathrm{off}}^{\mathrm{NN}} m_{\mathrm{NN}}
    \label{eq:R_adhesion_doublet}
\end{align}
with indices $\{1,2\}$ labeling the two droplets. The flux coupling bulk and surface densities is $j_\mathrm{NN} = \mathcal{R}_{m_\mathrm{NN}}$, and the tension at the droplet-droplet interface in the dilute limit $k_{\mathrm{on}}^{\mathrm{NN}} (k_{\rm{p}}^{\rm{N}}/k_{\rm{d}}^{\rm{N}})^2 \ll k_{\mathrm{off}}^{\mathrm{NN}} $ is
\begin{equation}
    \gamma_{\mathrm{c}} = \gamma_0 -\epsilon   \frac{
    k_{\mathrm{on}}^{\mathrm{NN}}}{k_{\mathrm{off}}^{\mathrm{NN}}}\left(\frac{k_{\rm{p}}^{\rm{N}}}{k_{\rm{d}}^{\rm{N}}}\right)^2 m_\mathrm{NN}^{\mathrm{max}}.
    \label{eq:contact_tension_doublet}
\end{equation}
Indeed, the force necessary to separate two adhesive cells has been shown to scale linearly with the squared total number of adhesion molecules \cite{maitre2011role, chu2004force}. 
In general, the kinetic rates of adhesion molecules 
can differ between contacting cells, subject to internal regulatory mechanisms. In Sec.~\ref{sec:signal_dependent_mechanics}, we analyse the case in which the production rate of adhesion molecules $k_\mathrm{p}^\mathrm{N}$ depends on a cell-intrinsic signaling state.

\subsection{Biochemical signaling interactions at contact surfaces}
\label{sec:contact-based_signaling} assume that the concentration of signal molecules determines the production rate of a regulator molecule (U) in the bulk (e.g. a cellular transcription factor) that controls in turn the production rates of adhesion and signaling molecules (Sec.~\ref{sec:coarse_grained_feedback}).[Chap.~15])

\subsubsection{Biochemical interactions between a cell and a signal-transmitting substrate}
\label{sec:singlet_signaling} 
We begin by considering a single cell in contact with a solid substrate that is functionalized with immobile ligands at a fixed uniform density $m_\mathrm{L}^\mathrm{max}$, similar to experimental systems developed for the Notch pathway in \emph{in vitro} assays \cite{beckstead2009methods} [Fig.~\ref{fig:1}(e)]. The cell contains receptor molecules, signaling molecules, and regulator molecules with bulk concentrations $c_\mathrm{R}, c_\mathrm{S},$ and $c_\mathrm{U}$ respectively, whose dynamics are coupled via the reactions at the contact surface. We do not explicitly consider a bulk concentration of ligands, because the substrate has no receptor molecules to bind to---the cell is only receiving, but not sending signals. To describe the signaling dynamics at the surface, we use Eq.~\eqref{eq:continuity_surface} for the surface densities of receptors $m_{\rm{R}}$, substrate-bound ligands $m_{\rm{L}}$ and receptor-ligand complexes $m_{\rm{RL}}$ with the reaction terms adapted from Khait et al. (2016) \cite{khait2016quantitative}
\begin{align}
    \mathcal{R}_{m_\mathrm{R}} &=
    k_{\rm{on}}^{\rm{R}} c_{\rm{R}}
    - (k_{\rm{off}}^{\rm{R}} + k_+ m_{\rm{L}}) m_{\rm{R}}
    + k_- m_{\rm{RL}}, \label{eq:reactions_receptors_singlet}\\
    \mathcal{R}_{m_\mathrm{L}} &=
    (k_- + k_{\rm{s}}) m_{\rm{RL}}
    - k_+ m_{\rm{L}} m_{\rm{R}},
    \label{eq:reactions_ligands_singlet}\\
    \mathcal{R}_{m_\mathrm{RL}} &= 
    k_+ m_{\rm{L}}m_{\rm{R}} 
    - (k_- + k_{\rm{s}}) m_{\rm{RL}}, \label{eq:reactions_complexes_singlet}
\end{align}
which are explained in the following [Fig.~\ref{fig:1}(e)]. Receptors are recruited to the surface with a rate set by $k^{\rm{R}}_{\rm{on}}$ (exocytosis) and they are removed from the surface with rate $k^{\rm{R}}_{\rm{off}}$ (endocytosis). Receptors at the contact surface bind ligands at a rate determined by $k_+$ to form receptor-ligand complexes, which unbind with rate $k_-$. Receptor-ligand complexes undergo an irreversible enzymatic cleavage with rate $k_{\rm{s}}$ upon which a fragment of the bound receptor molecule is released into the bulk and acts as a signaling molecule (S), the remaining part is degraded, and the ligand is released within the surface where it can bind to a new receptor molecule. The bulk concentration $c_\mathrm{S}$ of signaling molecules controls the bulk production rate $k_\mathrm{p}^\mathrm{U}(c_\mathrm{S})$ of the regulator U [see below, Eq.~\eqref{eq:kpU}]. The bulk concentrations of receptors $c_\mathrm{R}$ and signaling molecules $c_\mathrm{S}$ are coupled to the signaling dynamics at the contact via Eq.~\eqref{eq:bulk_concentration_average_Ac}.

We consider the ligands on the substrate to be fixed in place, such that $D_{m_\mathrm{L}}=0, D_{m_\mathrm{RL}}=0$ in Eq.~\eqref{eq:continuity_surface}. The density of unbound ligands is the difference between the total density of ligands covering the substrate and the density of receptor-ligand complexes $m_\mathrm{L} = m_\mathrm{L}^\mathrm{max} -  m_\mathrm{RL}$. Equations~\eqref{eq:continuity_surface} and \eqref{eq:reactions_complexes_singlet} together with this relation permit expressing the steady state concentration of receptor-ligand complexes in terms of the steady state receptor concentration as 
\begin{align}
    m_{\mathrm{RL}} &= \frac{
     m_\mathrm{R} 
    }{
    m_\mathrm{R}  + \dfrac{ k_{\mathrm{s}} + k_- }{ k_+}
    } m_{\mathrm{L}}^{\mathrm{max}} . \label{eq:steady_state_complexes_singlet}
\end{align}

Given Eqs.~\eqref{eq:continuity_surface}, \eqref{eq:reactions_receptors_singlet} and \eqref{eq:steady_state_complexes_singlet}, the  steady state relation for the distribution of receptors reads
\begin{equation}
        0 = D_{m_\mathrm{R}} \nabla^2 m_{\rm{R}}
        + k_{\rm{on}}^{\rm{R}} c_\mathrm{R} 
        - m_{\rm{R}} \left( k_{\rm{off}}^{\rm{R}}
         + \frac{k_s k_+ m_{\mathrm{L}}^{\mathrm{max}}
    }{
    k_+ m_\mathrm{R}  +  k_{\mathrm{s}} + k_-
    }  
    \right) .
        \label{eq:receptor_steady_state_singlet}
\end{equation}

In the limit where the rate of receptors binding to ligands on the substrate is large compared to the transport of receptors from the surface into the bulk, i.e. 
\begin{align}
    k_+ m_{\mathrm{L}}^{\mathrm{max}}
    \gg
    k_{\mathrm{off}}^{\mathrm{R}},
    \label{eq:limit1}
\end{align}
solutions of Eq.~\eqref{eq:receptor_steady_state_singlet} are uniform and follow
\begin{align}
    m_{\rm{R}} = \frac{
    k_{\rm{on}}^{\rm{R}} c_{\rm{R}} (k_- + k_{\rm{s}})
    }{
    k_+ (k_{\rm{s}} m_{\rm{L}}^{\rm{max}} - k_{\rm{on}}^{\rm{R}} c_{\rm{R}})
    },
    \label{eq:mR_cR_relation}
\end{align}
under boundary condition Eq.~\eqref{eq:bc_Ac}. The bulk and surface densities of receptors are coupled via the flux [Eq.~\eqref{eq:bulk_concentration_average_Ac}]
\begin{align}
    j_{\mathrm{R}} &= k_{\mathrm{on}}^{\mathrm{R}} c_{\mathrm{R}} - k_{\mathrm{off}}^{\mathrm{R}} m_{\mathrm{R}}.
    \label{eq:jR}
\end{align}
Using Eq.~\eqref{eq:mR_cR_relation} and assuming~\eqref{eq:limit1}, the flux can be approximated as $j_{\rm{R}} = k_{\rm{on}}^{\rm{R}} c_{\rm{R}}$
and the steady state bulk and surface densities of receptors that follow from Eqs.~\eqref{eq:bulk_concentration_average_Ac} and \eqref{eq:mR_cR_relation} are given by
\begin{align}
    c_{\rm{R}} &=
    \frac{
        k_{\rm{p}}^{\rm{R}} V   
    }{
        k_{\rm{d}}^{\rm{R}} V + k_{\rm{on}}^{\rm{R}} A_{\rm{c}}
    }, \\
    m_{\rm{R}} &=
    \frac{
        k_{\rm{on}}^{\rm{R}} k_{\rm{p}}^{\rm{R}} (k_- + k_{\rm{s}}) V
    }{
    k_+ [k_{\rm{s}} m_{\rm{L}}^{\rm{max}} (k_{\rm{on}}^{\rm{R}} A_{\rm{c}} + k_{\rm{d}}^{\rm{R}} V) - k_{\rm{on}}^{\rm{R}} k_{\rm{p}}^{\rm{R}} V]
    }. \label{eq:mR_solution_singlet}
\end{align}
For Notch receptors, reported values are $k_+ = \SI{0.167}{\micro\metre\squared\per\second}$ and $k_\mathrm{off}^{\rm{R}} = \SI{0.02}{\per\second}$ \cite{khait2016quantitative} [Tab.] and Notch activation assays with cells on ligand-coated substrates are performed with surface densities of up to $m_{\rm{L}}^{\rm{max}}\approx\SI{e5}{\per\micro\metre\squared}$ \cite{beckstead2009methods}, which justifies limit \eqref{eq:limit1} and allows to neglect the $k_{\rm{off}}^{\rm{R}}$-term in Eq.~\eqref{eq:receptor_steady_state_singlet}. Importantly, $m_{\rm{R}}$ and $c_{\rm{R}}$ have upper bounds: in the absence of ligands ($m_\mathrm{L}^\mathrm{max} = 0$), the steady state receptor density is uniform at $m_{\rm{R}}^0=k_\mathrm{on}^\mathrm{R} c_\mathrm{R}^0/k_\mathrm{off}^\mathrm{R}$ with bulk concentration $c_{\rm{R}}^0 = k_\mathrm{p}^\mathrm{R}/k_\mathrm{d}^\mathrm{R}$. Because receptors are removed upon receptor-ligand binding and subsequent cleavage, $m_{\rm{R}}^{0}$ and $c_\mathrm{R}^0$ are upper bounds to the steady state concentrations. If we estimate the term in brackets of Eq.~\eqref{eq:receptor_steady_state_singlet} using $m_{\rm{R}}=m_{\rm{R}}^0$ and typical parameter values as listed in Tab.~\ref{tab:parameter_values} (Appendix), neglecting $k_{\rm{off}}^{\rm{R}}$ is valid if $m_{\rm{L}}^{\rm{max}} \gg \SI{10}{\per\micro\metre\squared}$.

Given Eqs.~\eqref{eq:steady_state_complexes_singlet} and \eqref{eq:mR_solution_singlet}, the steady state density of receptor-ligand complexes is
\begin{align}
    m_{\rm{RL}} = \frac{
     k_{\rm{on}}^{\rm{R}} k_{\rm{p}}^{\rm{R}} V
    }{
    k_{\rm{s}} (k_{\rm{on}}^{\rm{R}} A_{\rm{c}} + k_{\rm{d}}^{\rm{R}} V)
    }.
\end{align}
The bulk concentration of signaling molecules follows Eq.~\eqref{eq:bulk_concentration_average_Ac}, without a bulk production term ($k_\mathrm{p}^\mathrm{S}=0$) and with flux $j_\mathrm{S}=-k_\mathrm{s}m_\mathrm{RL}$, arising from the cleavage of receptor-ligand molecules at the surface. The steady state bulk concentration is given by
\begin{align}
   c_\mathrm{S} = \frac{
   k_{\rm{on}}^{\rm{R}} k_{\rm{p}}^{\rm{R}} A_\mathrm{c}
   }{
   k_{\rm{d}}^{\rm{S}}( k_{\rm{on}}^{\rm{R}} A_{\rm{c}} + k_{\rm{d}}^{\rm{R}} V) 
   }.
    \label{eq:cs_singlet}
\end{align}
This relation shows how the received signal depends on the receptor-ligand kinetics, the volume, and the geometry of the adherent cell.

\subsubsection{Signaling interactions between contacting cells}\label{sec:signaling_doublet}
Next, we consider receptor-ligand interactions at the interface  between two cells indexed with $i,j\in\{1,2\}$ [Fig.~\ref{fig:1}(f)]. In addition to containing signaling and regulator molecules, each cell produces receptors as well as ligands, which exchange between bulk and surface---ligands are not substrate-bound with fixed positions as we considered in the preceding part. The receptors on the surface of cell $i$ bind to the ligands on the surface of the other cell $j$ and vice versa, producing respectively oriented receptor-ligand complexes. Upon cleavage they release signal molecules into the receptor-carrying cell $i$. Contrary to the way we treated substrate-bound ligands in the preceding section, ligands at the droplet interface are not released after cleavage of the receptor-ligand complexes, but are degraded together with the remaining receptor fragment instead \cite{seib2021role}. While some literature suggests that ligands can also be recycled after a signaling event or enter alternative signaling pathways \cite{vazquez2022reversible, kandachar2012endocytosis}, we here consider that receptors and ligands are always degraded after cleavage. Signaling molecules control the production of regulator molecules as before, which feed back onto the production terms. In line with the typical molecular mechanisms in Notch signaling, we consider an active regulation of \emph{ligand} production \cite{bray2016notch}.
As explained in Sec.~\ref{sec:microscopic_dynamics_general}, the steady state bulk concentrations of receptor, ligand, signaling, and regulator molecules are uniform within each cell with a value set by the flux balance condition [Eq.~\eqref{eq:bulk_concentration_average_Ac}].
To capture the reaction-diffusion dynamics at the interface, we use Eq.~\eqref{eq:continuity_surface} for receptors, ligands, and complexes using the reaction terms \cite{khait2016quantitative}
\begin{align}
    \mathcal{R}_{\mathrm{R}} &=
    k_{\mathrm{on}}^{\mathrm{R}} c_{\mathrm{R},i}
    - (k_{\mathrm{off}}^{\mathrm{R}} + k_+ m_{\mathrm{L},j}) m_{\mathrm{R},i}
    + k_- m_{\mathrm{RL},i} \label{eq:reactions_receptors_doublet},\\
    \mathcal{R}_{\mathrm{L}} &=
    k_{\mathrm{on}}^L c_{\mathrm{L},j}
    - (k_{\mathrm{off}}^{\mathrm{L}} + k_+ m_{\mathrm{R},i}) m_{\mathrm{L},j}
    + k_- m_{\mathrm{RL},i} \label{eq:reactions_ligands_doublet}, \\
    \mathcal{R}_{\mathrm{RL}} &= 
    k_+ m_{\mathrm{R},i}m_{\mathrm{L},j} 
    - (k_- + k_{\mathrm{s}}) m_{\mathrm{RL},i}, \label{eq:reactions_complexes_doublet}
\end{align}
with rate constants as described in Sec.~\ref{sec:singlet_signaling}. Under boundary condition Eq.~\eqref{eq:bc_Ac}, steady state solutions of Eq.~\eqref{eq:continuity_surface} for the densities of receptors, ligands and receptor-ligand complexes with the reaction terms Eqs.~\eqref{eq:reactions_receptors_doublet}-\eqref{eq:reactions_complexes_doublet} are uniform and follow the relations
\begin{align}
    m_{\mathrm{R},i} &= \frac{
    k_{\rm{on}}^{\rm{R}} c_{\mathrm{R},i} (k_{\rm{s}} + k_- )
    }{
    k_{\rm{off}}^{\rm{R}}(k_{\rm{s}}+k_-) + k_+ k_{\rm{s}} m_{\mathrm{L},j}
    }, \label{eq:relation_mR_doublet}\\
    m_{\mathrm{L},j} &= \frac{
    k_{\rm{on}}^{\rm{L}} c_{\mathrm{L},j} (k_{\rm{s}} + k_- )
    }{
    k_{\rm{off}}^{\rm{L}}(k_{\rm{s}}+k_-) + k_+ k_{\rm{s}} m_{\mathrm{R},i}
    }. \label{eq:relation_mL_doublet}
\end{align}
Ligand molecules are only produced in the bulk, but not at the surface, thus, the steady state concentrations of bulk and surface densities have the upper limits $c_{\rm{L}}^{\mathrm{0}}=k_{\mathrm{p}}^{\rm{L}}/k_{\mathrm{d}}^{\rm{L}}$ and $m_{\rm{L}}^{\mathrm{0}}=k_{\mathrm{on}}^{\rm{L}}c^{\mathrm{0}}/k_{\mathrm{off}}^{\rm{L}}$. Together with Eqs.~\eqref{eq:relation_mR_doublet} and \eqref{eq:bulk_concentration_average_Ac} for the receptor bulk concentration, one can define a lower limit for the surface density of receptors
\begin{align}
    m_{\rm{R}}^{\rm{min}} = \frac{
    k_{\rm{on}}^{\rm{R}} k_{\rm{off}}^{\rm{L}} (k_- + k_{\rm{s}}) k_{\rm{p}}^{\rm{R}} k_{\rm{d}}^{\rm{L}} V
    }{
    k_{\rm{on}}^{\rm{L}} k_+ k_{\rm{s}} k_{\rm{p}}^{\rm{L}} (k_{\rm{on}}^{\rm{R}} A_{\rm{c}} + k_{\rm{d}}^{\rm{R}}V) + k_{\rm{d}}^{\rm{L}} k_{\rm{off}}^{\rm{R}}(k_- + k_{\rm{s}}) k_{\rm{d}}^{\rm{R}}V
    }. \label{eq:mRmin}
\end{align}
In line with Khait et al. 2016 \cite{khait2016quantitative}, we consider that cells produce an excess of receptors compared to the number of ligands, i.e. $k_{\rm{p}}^{\rm{R}} \gg k_{\rm{p}}^{\rm{L}}$, and similar to limit~\eqref{eq:limit1} for the single cell we assume that the endocytosis rate of ligands is small compared to the rate of binding
\begin{align}
    k_+ m_{\mathrm{R}}^{\rm{min}} \gg k_{\rm{off}}^{\rm{L}},
    \label{eq:limit2}
\end{align}
allowing to neglect the $k_{\rm{off}}^{\rm{L}}$-term in Eq.~\eqref{eq:reactions_ligands_doublet}. The bulk and surface densities of ligands are coupled in Eq.~\eqref{eq:bulk_concentration_average_Ac} via the flux 
\begin{align}
j_{\mathrm{L},j} = k_{\mathrm{on}}^{\mathrm{L}} c_{\mathrm{L},j} - k_{\mathrm{off}}^{\mathrm{L}} m_{\mathrm{L},j},
\end{align}
which using Eq.~\eqref{eq:relation_mL_doublet} and assuming~\eqref{eq:limit2} can be written as $j_{\rm{L},j} = 
    k_{\rm{on}}^{\rm{L}} c_{\rm{L},j}$.
In this limit, solving Eqs.~\eqref{eq:continuity_surface} with the reaction terms Eqs.~\eqref{eq:reactions_receptors_doublet}--\eqref{eq:reactions_complexes_doublet} under boundary condition Eq.~\eqref{eq:bc_Ac} together with Eq.~\eqref{eq:bulk_concentration_average_Ac} and boundary flux Eq.~\eqref{eq:jR} for the bulk density of receptors yields
\begin{widetext}
\begin{align}
    c_{\mathrm{R},i} &= \dfrac{
    k_{\rm{p}}^{\rm{R}} ( k_{\rm{on}}^{\rm{L}} A_{\rm{c}} + k_{\rm{d}}^{\rm{L}} V) 
        - k_{\rm{p}}^{\rm{L}} k_{\rm{on}}^{\rm{L}} A_{\rm{c}}
    }{
    k_{\rm{d}}^{\rm{R}} ( k_{\rm{on}}^{\rm{L}} A_{\rm{c}} + k_{\rm{d}}^{\rm{L}} V)
    }, \\
     c_{\mathrm{L},j} &= \frac{
    k_{\rm{p}}^{\rm{L}} V
    }{
    k_{\rm{on}}^{\rm{L}} A_{\rm{c}} + k_{\rm{d}}^{\rm{L}} V
    }, \\
    m_{\mathrm{R},i} &= \frac{
    A_{\rm{c}} k_{\rm{on}}^{\rm{L}} k_{\rm{on}}^{\rm{R}} (k_{\rm{p}}^{\rm{R}} - k_{\rm{p}}^{\rm{L}}) + ( k_{\rm{d}}^{\rm{L}} k_{\rm{on}}^{\rm{R}} k_{\rm{p}}^{\rm{R}} - k_{\rm{d}}^{\rm{R}} k_{\rm{on}}^{\rm{L}} k_{\rm{p}}^{\rm{L}})V
    }{
    k_{\rm{d}}^{\rm{R}} k_{\rm{off}}^{\rm{R}} (k_{\rm{on}}^{\rm{L}} A_{\rm{c}} + k_{\rm{d}}^{\rm{L}} V)
    }, \\
    m_{\mathrm{L},j} &= \frac{
    k_{\rm{on}}^{\rm{L}} k_{\rm{p}}^{\rm{L}} k_{\rm{d}}^{\rm{R}} k_{\rm{off}}^{\rm{R}} (k_- + 
   k_{\rm{s}}) V
    }{
    k_+ k_{\rm{s}} (A_{\rm{c}} k_{\rm{on}}^{\rm{L}} k_{\rm{on}}^{\rm{R}} (k_{\rm{p}}^{\rm{R}} - k_{\rm{p}}^{\rm{L}}) + ( k_{\rm{d}}^{\rm{L}} k_{\rm{on}}^{\rm{R}} k_{\rm{p}}^{\rm{R}} - k_{\rm{d}}^{\rm{R}} k_{\rm{on}}^{\rm{L}} k_{\rm{p}}^{\rm{L}})V)
    }, \\
    m_{\mathrm{RL},i} &= \frac{
    k_{\rm{on}}^{\rm{L}} k_{\rm{p}}^{\rm{L}} V
    }{
    k_{\rm{s}}(k_{\rm{on}}^{\rm{L}} A_{\rm{c}} + k_{\rm{d}}^{\rm{L}} V)
    }.
\end{align}
\end{widetext}
The steady state bulk concentration of signaling molecules following Eq.~\eqref{eq:bulk_concentration_average_Ac} with $j_\mathrm{S}=-k_\mathrm{s}m_\mathrm{RL}$ and $k_\mathrm{p}^\mathrm{S}=0$ as before is
\begin{align}
   c_{\mathrm{S},i} = \frac{
    k_{\rm{on}}^{\rm{L}} k_{\rm{p}}^{\rm{L}} A_{\rm{c}}
    }{
    k_{\rm{d}}^{\rm{S}} (k_{\rm{on}}^{\rm{L}} A_{\rm{c}} + k_{\rm{d}}^{\rm{L}} V)
    }.
    \label{eq:cs_doublet}
\end{align}
As Eq.~\eqref{eq:cs_singlet}, this relation shows how the signal exchanged between interacting cells depends on the receptor-ligand kinetics and the geometry of the cells.

\section{Mechanochemical feedback between adhesion and cell-cell signaling}
\label{sec:coarse_grained_feedback}

\begin{figure}
    \centering
    \includegraphics[width=\linewidth]{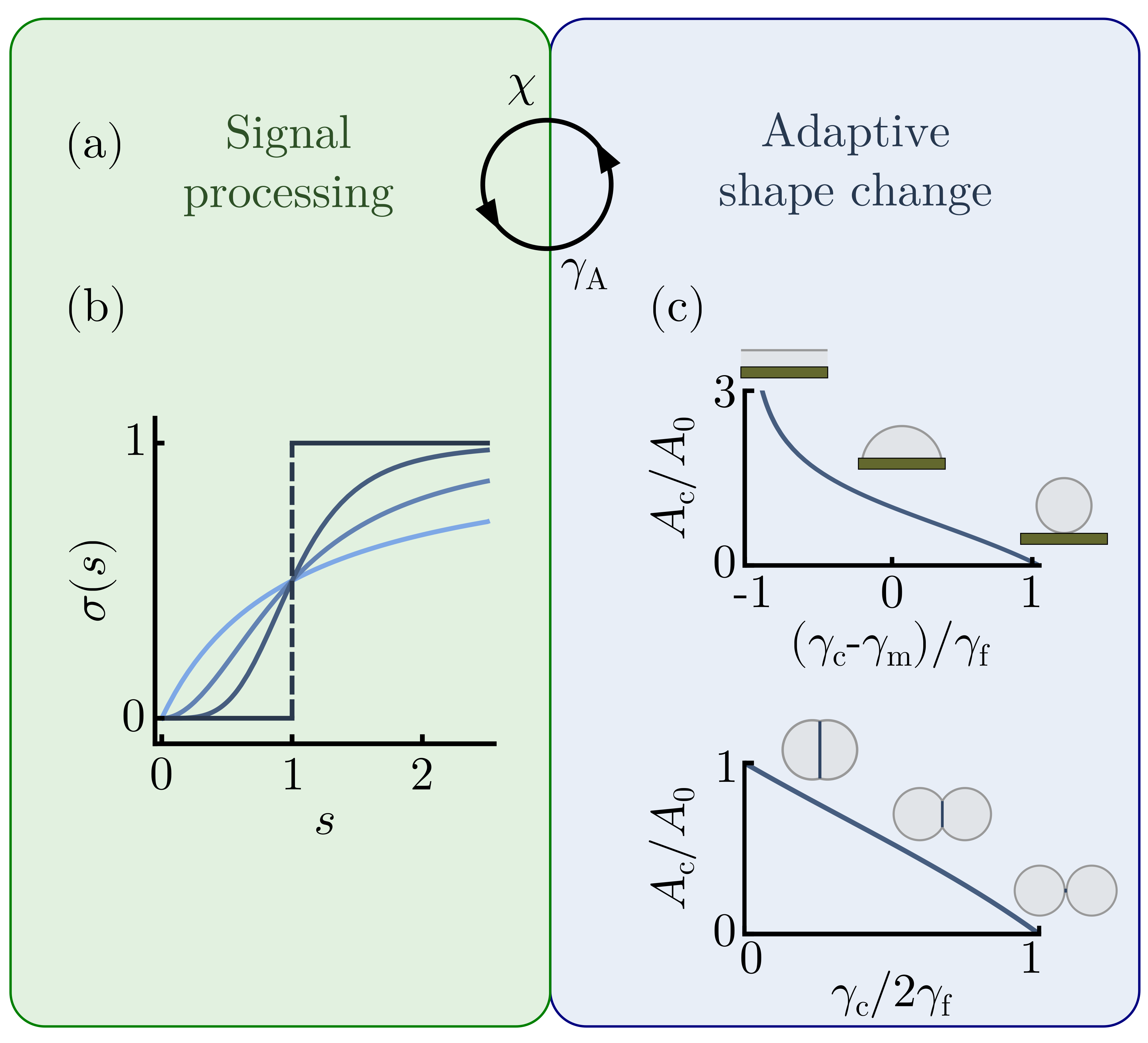}
    \caption{(a) The feedback between signal processing and active mechanics olled by the \emph{signal susceptibility} parameter $\chi$ and the \emph{adaptive adhesion coefficient} $\gamma_{\rm{A}}$. (b) Signals $s$ are processed with a sigmoidal response function Eq.\,\eqref{eq:sigma}, which arises from the interaction between signaling molecules and regulator molecules [Eq.~\eqref{eq:signal_definition_singlet},\eqref{eq:signal_definition_doublet}]. In the limit $h\rightarrow\pm\infty$, the response is a step function.
    (c) The equilibrium contact area of an incompressible droplet with a substrate (top) and another droplet (bottom) is determined by the tension ratio between contact and free surface [Eq.~\eqref{eq:contact_area}][Fig.~\ref{fig:1}(a-b)].
    }
    \label{fig:2}
\end{figure}

In response to external signals, cells typically change their gene expression through transcription factors and transcriptional regulators, thereby controlling the production rates of diverse proteins, including adhesion and signaling molecules \cite[Chap.~7]{alberts2022molecular}. Responding to shape-dependent signals, and feeding back onto both mechanics and signaling, these internal regulatory states couple the processes described in Sections Secs.~\ref{sec:adhesion} and \ref{sec:contact-based_signaling}. 

In the following, we derive the macroscopic equations that govern the evolution of such internal states, taking into account how the received signals depend on contact geometry, and how the contact geometry in turn is set by adaptive adhesion. In particular, from the microscopic kinetics of adhesion and signaling molecules (Secs.\ref{sec:adhesion},\ref{sec:contact-based_signaling}) we obtain two macroscopic feedback parameters: the \emph{signal susceptibility} determines how the contact area affects the magnitude of transmitted signals, and the \emph{adaptive adhesion coefficient} controls how the received signals feed back onto the contact mechanics [Fig.~\ref{fig:2}(a)]. 

\subsection{Evolution of a macroscopic signaling state}

We introduce a macroscopic internal cell state variable that responds to area-dependent biochemical signals defined by the uniform bulk concentration of a regulator molecule $U$--representing for instance a transcription factor.
The bulk concentration $c_{\rm{U}}$ is governed by Eq.~\eqref{eq:bulk_average_dynamic} with $j_{\mathrm{U}}=0$, and we assume that the effective production rate of regulator molecules depends on the steady state concentration of signal molecules $k_\mathrm{p}^\mathrm{U}(c_\mathrm{s})$---the more signal molecules are present, the more regulator molecules are produced [Fig.~\ref{fig:2}(b)]. 
The regulation of genes and the synthesis of new proteins involves multiple steps and molecular intermediates, which leads to the presence of nonlinear effects like cooperative binding and multimerization, commonly captured using Hill functions \cite{santillan2008use}. Similar to previous studies modeling canonical Notch signaling \cite{erzberger2020mechanochemical, binshtok2018modeling, corson2017self, collier1996pattern}, we therefore assume that steady state concentrations of  $U$ are bounded within a range $c_{\mathrm{U}}^{\mathrm{min}} \leq c_{\mathrm{U}} \leq c_{\mathrm{U}}^{\mathrm{max}}$ and we consider a nonlinear production rate with Hill coefficient $h$
\begin{align}
    k_{\mathrm{p}}^{\mathrm{U}}(c_{\mathrm{S}}) = \frac{1}{\tau_\mathrm{u}}\left(
    c_{\mathrm{U}}^{\mathrm{min}} + \frac{
    (c_{\mathrm{U}}^{\mathrm{max}} - c_{\mathrm{U}}^{\mathrm{min}})
    }{
    1 + \left(\frac{c_{\mathrm{S}}^{\mathrm{crit}}}{c_{\mathrm{S}}}\right)^h
    }
    \right),
    \label{eq:kpU}
\end{align}
in which $\tau_{\mathrm{u}} = 1/k_\mathrm{d}^\mathrm{U}$ is the characteristic time scale on which $c_\mathrm{U}$ is changing, and $c_{\mathrm{S}}^{\mathrm{crit}}$ is the critical concentration at the inflection point.
The saturating response to the received signal [Eq.~\eqref{eq:kpU}] permits introducing a dimensionless \emph{signaling state} variable 
\begin{align}
u := \frac{
c_{\mathrm{U}} - c_{\mathrm{U}}^{\mathrm{min}}
}{
c_{\mathrm{U}}^{\mathrm{max}}-c_{\mathrm{U}}^{\mathrm{min}}
},
\label{eq:definition_u}
\end{align} 
normalized to the response range such that $u \in [0,1]$. Eqs.~\eqref{eq:bulk_average_dynamic}, \eqref{eq:kpU}, and \eqref{eq:definition_u} together with the definition of a normalized received signal
\begin{align}
    s = c_\mathrm{S}/c_{\mathrm{S}}^{\mathrm{crit}}
    \label{eq:signal_def}
\end{align} 
lead to a dynamical equation for the evolution of the internal state
\begin{align}
    \tau_{\mathrm{u}} \frac{du}{dt} = \sigma(s) - u,
    \label{eq:u_ode}
\end{align}
with sigmoidal response function [Fig.~\ref{fig:2}(b)]
\begin{align}
    \sigma(s) = \frac
        { s^h }
        { 1 + s^h }.
         \label{eq:sigma}
\end{align}
Given that the regulation of protein concentrations through transcriptional changes requires tens of minutes to hours and can vary greatly between different protein species \cite{SHAMIR20161302, rusilowicz2022protein}, we assume that the timescale associated with the regulator turnover $\tau_\mathrm{u}$ dominates the dynamics of the system. On this timescale, we assume that bulk and surface concentrations relax to their steady state solutions. In cells, concentration and shape dynamics are indeed typically at least an order of magnitude faster---set by diffusive, biochemical, and viscoelastic timescales which are on the order of seconds to minutes \cite{wyatt2016question, khait2016quantitative, sankaran2009diffusion, SHAMIR20161302}.
In the following sections, we discuss how the internal state dynamics govern the production of adhesion and ligand molecules.

\subsection{Signal susceptibility}

In general, the received signal \eqref{eq:signal_def} depends non-linearly on the size of the contact area, i.e. $s(A_{\rm{c}})$ according to \eqref{eq:cs_singlet} or \eqref{eq:cs_doublet}. When the number of receptors that are recruited to the surface and lost in the signaling process is small compared to the turnover of molecules in the bulk, we can expand the bulk concentration of signal molecules [Eq.~\eqref{eq:cs_singlet},\eqref{eq:cs_doublet}], and obtain a relation that is linear in the contact area. In the limit in which receptors interact with an excess of ligands [Eq.~\eqref{eq:limit1}], e.g. 
for the single cell on the functionalized substrate, the expression reads
\begin{align}
    c_{\mathrm{S}} = \frac{
   k_{\rm{p}}^{\rm{R}} k_{\rm{on}}^{\rm{R}} A_{\rm{c}}
   }{
   k_{\rm{d}}^{\rm{S}} k_{\rm{d}}^{\rm{R}} V
   }
    + \mathcal{O}\left( \left( \frac{
        k_\mathrm{on}^\mathrm{R} A_\mathrm{c}
        }{
        k_\mathrm{d}^\mathrm{R} V 
        } \right)^2 \right).
    \label{eq:cs_expansion_singlet}
\end{align}
Indeed, \emph{in vitro} experiments revealed a roughly linear relation between the Notch signaling response and the contact area, including for large contacts~\cite{khait2016quantitative}. The received signal [Eq.~\ref{eq:signal_def}] 
can then be written as
\begin{align}
    s = \chi \frac{A_\mathrm{c}}{A_0},
    \label{eq:signal_definition_singlet}
\end{align}
in which we introduced the \emph{signal susceptibility}
\begin{align}
    \chi = \frac{
    k_\mathrm{p}^{\rm{R}} k_{\rm{on}}^{\rm{R}}
    }{
    c_\mathrm{S}^\mathrm{crit} k_\mathrm{d}^\mathrm{S} k_{\rm{d}}^{\rm{R}} V
    } A_\mathrm{0}
    \label{eq:chi_singlet}
\end{align}
using the definition of the volume-dependent reference area $A_0 = (3V/2)^{2/3} \pi^{1/3}$. The volume-dependence of the susceptibility arises because the degradation of molecules in the bulk scales with the volume, and due to the reference area $A_0$, yielding a scaling of $\chi\propto V^{-1/3}$. However, in cells where protein degradation does not increase with the cell volume, the signal susceptibility might increase with volume. We can estimate the order of magnitude of the susceptibility [eq.~\ref{eq:chi_singlet}] using $k_{\rm{on}}^{\rm{R}} k_{\rm{p}}^{\rm{R}}/k_{\rm{d}}^{\rm{R}}=\SI{2}{\per\micro\metre\squared\per\second}$ \cite{khait2016quantitative}, $V=\SI{500}{\cubic\micro\metre}$, $k_{\rm{d}}^{\rm{S}}=\SI{5e-3}{\per\minute}$ \cite{ilagan2011real, fryer2004mastermind} and $c_{\rm{S}}^{\rm{crit}}=1000/V$ \cite{tveriakhina2024temporal} yielding $\chi \sim 3000$.

\subsubsection{Signal-dependent production of ligands}

Mutually inhibitory Notch signals typically lead to a decrease in the production rate of ligands in response to received signals \cite{sjoqvist2019say,bray2016notch}. We therefore consider that the production rate of ligands is a monotonously \emph{decreasing} function of the regulator concentration $c_{\mathrm{U}}$. We assume that no ligands are produced at $c_{\mathrm{U}} = c_{\mathrm{U}}^{\mathrm{max}}$, i.e. $k_{\mathrm{p}}^{\mathrm{L}}(c_\mathrm{U}^{\mathrm{max}}) = 0$, and we expand $k_{\mathrm{p}}^{\mathrm{L}}$ to first order around $c_\mathrm{U}^{\rm{max}}$
\begin{align}
    k_{\mathrm{p}}^{\mathrm{L}}(c_{\mathrm{U}}) = 
    \left. \frac{d k_{\mathrm{p}}^{\mathrm{L}}}{d c_\mathrm{U}} \right|_{c_\mathrm{U}^\mathrm{max}}(c_{\mathrm{U}} - c_\mathrm{U}^\mathrm{max})
    + \mathcal{O}\left ((c_\mathrm{U} - c_\mathrm{U}^\mathrm{max})^2 \right).
\end{align}
Using the definition of $u$ [Eq.~\eqref{eq:definition_u}] it follows that 
\begin{align}
     k_{\mathrm{p}}^{\mathrm{L}}(u) =  \left(- \left. \frac{d k_{\mathrm{p}}^{\mathrm{L}}}{d c_\mathrm{U}} \right|_{c_\mathrm{U}^\mathrm{max}} \right) (c_{\mathrm{U}}^\mathrm{max} - c_\mathrm{U}^\mathrm{min}) (1-u).
     \label{eq:kpL_of_u}
\end{align}
Linearizing the bulk concentration of signal molecules [Eq.~\eqref{eq:cs_doublet}] as before, we obtain
\begin{align}
    c_{\mathrm{S},i} = \frac{
   k_{\rm{p}}^{\rm{L}} k_{\rm{on}}^{\rm{L}} A_{\rm{c}}
   }{
   k_{\rm{d}}^{\rm{S}} k_{\rm{d}}^{\rm{L}} V
   }
    + \mathcal{O}\left( \left( \frac{
        k_\mathrm{on}^\mathrm{L} A_\mathrm{c}
        }{
        k_\mathrm{d}^\mathrm{L} V 
        } \right)^2 \right),
    \label{eq:cs_expansion_doublet}
\end{align}
with which the signal $s_{ij}=c_{\mathrm{s},i}/c_\mathrm{s}^\mathrm{crit}$ received by cell $i$ from cell $j$ is
\begin{align}
    s_{ij} = \chi \frac{A_\mathrm{c}}{A_0} (1-u_j)
    \label{eq:signal_definition_doublet}
\end{align}
with the signal susceptibility in the ligand-limited case given by
\begin{align}
\chi = \frac{
    k_{\rm{on}}^{\rm{L}} A_\mathrm{0} (c_{\mathrm{U}}^\mathrm{max} - c_\mathrm{U}^\mathrm{min})
    }{
    c_\mathrm{S}^\mathrm{crit} k_\mathrm{d}^\mathrm{S} k_{\rm{d}}^{\rm{L}} V
    }\left(- \left. \frac{d k_{\mathrm{p}}^{\mathrm{L}}}{d c_\mathrm{U}} \right|_{c_\mathrm{U}^\mathrm{max}} \right)
    \label{eq:chi_doublet}
\end{align}
The expression of the susceptibility is similar to Eq.~\eqref{eq:chi_singlet}, but depends on the production, decay and transport rates of ligands rather than receptors. While Eq.~\eqref{eq:chi_singlet} holds when receptors interact with an excess of ligands [Eq.~\eqref{eq:limit1}], an excess of receptors compared to ligands [Eq.~\eqref{eq:limit2}] leads to Eq.~\eqref{eq:chi_doublet}. In the ligand-limited case, contributions to the susceptibility can be further distinguished based on properties of the signal \emph{sending} cell, specifically $k_{\rm{p}}^{\rm{L}}k_{\rm{on}}^{\rm{L}}/k_{\rm{d}}$, and properties of the signal \emph{receiving} cell, including $c_{\rm{s}}^{\rm{crit}}$ and $k_{\rm{d}}^{\rm{S}}$ \cite{PRLJoint}.

Interestingly, the signal susceptibility is independent of the cleavage rate $k_{\rm{s}}$ in both cases. A common experimental perturbation to Notch signaling is the pharmacological inhibition of the enzyme cleaving the receptor-ligand complexes (treatment of cells with $\gamma$-secretase inhibitors) \cite{olsauskas2013gamma}. Our result suggests that the signal susceptibility and thus the steady state concentration of signaling molecules is independent of $k_{\rm{s}}$ unless cleavage is completely prevented.

\subsection{Signal-dependent active mechanics}
\label{sec:signal_dependent_mechanics}

In many biological systems, for instance mechanosensory epithelia \cite{cohen2023precise, cohen2022mechano, erzberger2020mechanochemical} or synthetically engineered systems \cite{toda2018programming}, adhesion molecules are expressed downstream of contact-based signals. Accordingly, we consider that the production rate of adhesion molecules $k_{\mathrm{p}}^{\mathrm{N}}(c_{\mathrm{U}})$ is a monotonously increasing function of the regulator concentration [Fig.~\ref{fig:1}(c)]. 
We assume that $k_{\mathrm{p}}^{\mathrm{N}}$ vanishes for $c_{\mathrm{U}} \leq c_{\mathrm{U}}^{\mathrm{min}}$, i.e. no adhesion molecules are produced when the regulator concentration drops below a concentration $c_{\mathrm{U}}^{\mathrm{min}}$, and we linearize $k_{\mathrm{p}}^{\mathrm{N}}$ around $c_{\mathrm{U}}^\mathrm{min}$
\begin{align}
    k_{\mathrm{p}}^{\mathrm{N}}(c_\mathrm{U}) = 
    \left. \frac{
    d k_{\mathrm{p}}^{\mathrm{N}} 
    }{
    d c_\mathrm{U}
    }\right\rvert_{c_\mathrm{U}^\mathrm{min}} (c_\mathrm{U} - c_\mathrm{U}^\mathrm{min}) + 
    \mathcal{O}\left((c_\mathrm{U}-c_\mathrm{U}^\mathrm{min})^2\right).
    \label{eq:kpN_linear}
\end{align}
With Eq.~\eqref{eq:definition_u}, the surface tension at the contact site of a single cell with an underlying substrate [Eq.~\eqref{eq:contact_tension_singlet}] can then be written as
\begin{align}
    \gamma_{\mathrm{c}} = \gamma_0 - \gamma_{\mathrm{A}} u
    \label{eq:adaptive_adhesion_singlet}
\end{align}
where we define the \emph{adaptive adhesion coefficient}
\begin{align}
    \gamma_{\mathrm{A}} = \epsilon \frac{
    k_{\mathrm{on}}^{\mathrm{N}} (c_\mathrm{U}^\mathrm{max} - c_\mathrm{U}^\mathrm{min})
    }{
    k_{\mathrm{off}}^{\mathrm{N}} k_{\mathrm{d}}^\mathrm{N}
    } \left. \frac{
    d k_{\mathrm{p}}^{\mathrm{N}} 
    }{
    d c_\mathrm{U}
    }\right\rvert_{c_\mathrm{U}^\mathrm{min}} m_\mathrm{L}^\mathrm{max}.
\label{eq:adaptive_adhesion_coefficient_singlet}
\end{align}tension at the interface between two contacting cells  Eq.~\eqref{eq:contact_tension_doublet} is 
\begin{align}
    \gamma_{\mathrm{c}} = \gamma_0 - \gamma_{\mathrm{A}} u_1 u_2,
    \label{eq:adaptive_adhesion_doublet}
\end{align}
with
\begin{align}
    \gamma_{\mathrm{A}} = \epsilon \frac{
    k_{\mathrm{on}}^{\mathrm{NN}} (c_{\rm{U}}^{\rm{max}} - c_{\rm{U}}^{\rm{min}})^2
    }{
    k_{\mathrm{off}}^{\mathrm{NN}} (k_{\mathrm{d}}^{\rm{N}})^2
    } \left(\left. \frac{d k_{\mathrm{p}}^{\mathrm{N}}}{d c_\mathrm{U}} \right|_{c_\mathrm{U}^\mathrm{max}}\right)^2 m_\mathrm{NN}^\mathrm{max}.
\label{eq:adaptive_adhesion_coefficient_doublet}
\end{align}

Eqs.~\eqref{eq:adaptive_adhesion_singlet} and \eqref{eq:adaptive_adhesion_doublet} are identical except for the squared terms arising from the production and decay of adhesion molecules, because both cells need to contribute molecules for the formation of adhesion complexes at the interface [Fig.~\ref{fig:1}(d)].

The adaptive adhesion coefficient $\gamma_{\rm{A}}$ has units of energy per area. The tension at cellular surfaces is usually dominated by the active contractility of the actomyosin cortex and is on the order of \SIrange{0.05}{0.5}{\nano\newton\per\micro\metre} \cite{yamamoto2023dissecting, godard2020apical, chugh2017actin, maitre2016asymmetric, fischer2014quantification}. The tension at a cell-cell or cell-substrate interface can be inferred from the contact angle if the tension at the free surface $\gamma_{\rm{f}}$ is known \cite{roffay2021inferring}. For instance, for $\gamma_{\rm{f}}=\SI{0.1}{\nano\newton\per\micro\metre}$, a range of contact angles $\theta=$10-100$^{\circ}$ corresponds to interfacial tensions of  approx. \SIrange{0.13}{0.2}{\nano\newton\per\micro\metre}. 
 Combined fluorescence-based density measurements of the adhesion molecule E-cadherin in \emph{C.elegans} embryos suggests that changes in the interfacial tension $\gamma_{\rm{c}}$ due to expression of adhesion molecules---as described by the adaptive adhesion term---are up to $\SI{0.41}{\nano\newton\per\micro\metre}$ \cite{yamamoto2023dissecting}, demonstrating that regulation of adhesion molecule expression provides access to a large range of shape configurations. Assuming a lateral distance of $\sim\SI{10}{\nano\metre}$ between adhesion molecules \cite{fichtner2014covalent}, i.e. a surface density of $\SI{10,000}{\per\micro\metre\squared}$, the effective surface energy per adhesion molecule would be $\epsilon \approx 4\times10^{-17}~\text{J}$---several orders larger than $k_{\rm{B}}T$. Indeed, adhesion complexes in cells interact with different molecules and their formation depends also on anchoring to the cytoskeleton, which itself exhibits complex dynamics and feedback effects \cite{sitarska2020pay, maitre2012adhesion, maitre2013three, arslan2024adhesion}, thus $\epsilon$ corresponds to an effective energy per adhesion complex that captures more than just the binding energy between two adhesion molecules.

\subsection{Equilibrium shapes of cells with uniform interfacial tensions}

\label{sec:singlet_bistability}

Equations~\eqref{eq:signal_definition_singlet},\eqref{eq:signal_definition_doublet} and \eqref{eq:adaptive_adhesion_singlet},\eqref{eq:adaptive_adhesion_doublet} describe respectively how transmitted signals depend on the area of the cell-cell or cell-substrate interface, and how the interfacial tension in turn depends on the internal states. Given that the mechanochemical dynamics are dominated by the slowest timescale $\tau_{\rm{u}}$, set by transcriptional regulation, the contact areas across which signals are exchanged are determined quasi-instantaneously by the conjugate interfacial tension $\gamma_{\rm{c}}$. 
Neglecting any non-uniform contributions to the surface tensions, we assume that the cell shapes can be approximated by minimal surface configurations, i.e. that minimize the effective surface energy of $N$ incompressible droplets which form $n$ contacts
\begin{align}
    E = \sum_{i=1}^N \gamma_{\mathrm{f}}A_{\mathrm{f},i}
    + \sum_{j=1}^n \gamma_{\mathrm{c}} A_{\mathrm{c},j} 
    + \gamma_{\mathrm{c}} A_{\rm{m}},
    \label{eq:surface_energy}
\end{align}
in which $\gamma_{\rm{c}}$, $\gamma_{\rm{f}}$ and $\gamma_{\rm{m}}$ are the uniform surface tensions of the cell-cell or cell-substrate contact interfaces $A_{\mathrm{c},i}$, of the free surface areas $A_{\mathrm{f},i}$, and of the substrate-medium interface $A_{\rm{m}}$, respectively [Fig.~\ref{fig:1}(a,b)].  
In the minimal surface configuration, the droplets acquire a spherical-cap shape with contact area 
\begin{align}
    \frac{A_{\mathrm{c}}}{A_0} = 
    \left[ 1 - \cos\left( \theta/2 \right)^2 \right]
    \left[ 
        \frac{2}
        {(2 - \cos\left( \theta/2 \right))
        (1 + \cos\left( \theta/2 \right))^2} 
    \right]^{2/3}.
    \label{eq:contact_area}
\end{align}
in which $\theta$ is the contact angle with $\cos(\theta/2)=(\gamma_{\rm{c}}-\gamma_{\rm{m}}/\gamma_{\rm{f}})$ for a single adherent droplet, and $\cos(\theta/2)=(\gamma_{\rm{c}}/2\gamma_{\rm{f}})$ for a pair of droplets [Fig.~\ref{fig:2}(c,d)]. Indeed, biological cells have been found in minimal surface configurations in many contexts [Fig.~\ref{fig:1}(a-b)], including \cite{fabreges2024temporal,yamamoto2023dissecting,erzberger2020mechanochemical,godard2020apical,maitre2016asymmetric,maitre2012adhesion}.

\section{Macroscopic dynamics of mechanochemical droplets}
\label{sec:nonlinear_dynamics}

\begin{figure}
    \centering
    \includegraphics[width=\linewidth]{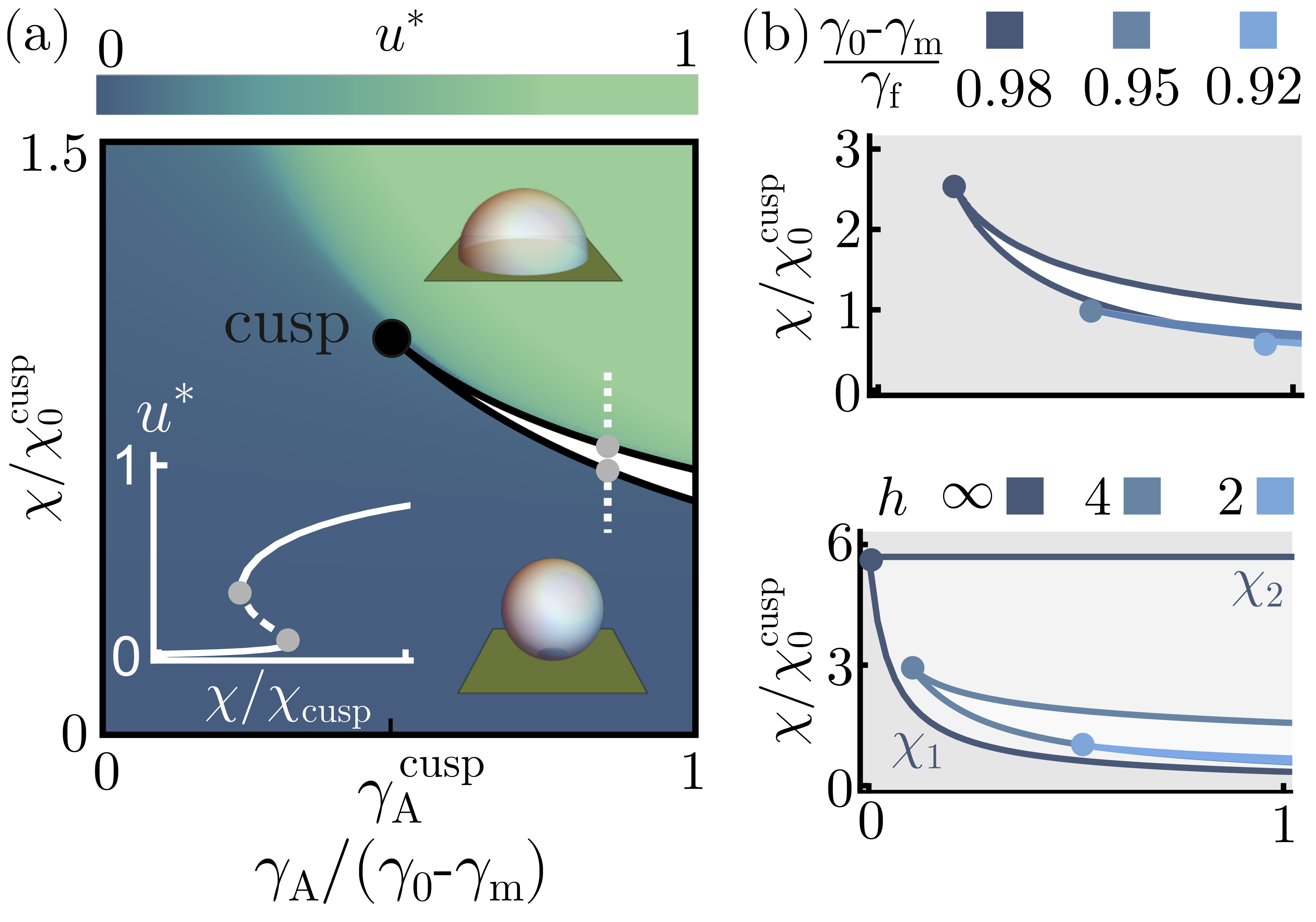}
    \caption{Adaptive adhesion leads to bistability. (a) The parameter diagram, derived via numerical continuation (Appendix~\ref{sec:numerical_methods}), contains a bistable regime (white) bounded by saddle-node bifurcations (black lines) converging in a codimension-2 cusp point, separating regimes of strong and weak substrate wetting 
    (b) The size of the bistable regime increases with the tension ratio $(\gamma_0-\gamma_{\rm{m}})/\gamma_{\rm{f}}$ (top) and with the Hill coefficient $h$ of the nonlinear response function [Eq.\eqref{eq:sigma}] (bottom) ($\chi_0^{\rm{cusp}}$: reference susceptibility at cusp for $(\gamma_0-\gamma_{\rm{m}})/\gamma_{\rm{f}}=0.95$, $h=2$).
    Parameter values for each diagram listed in Tab.~\ref{tab:parameter_values_figures}
    }
    \label{fig:3}
\end{figure}

\begin{figure}
    \centering
    \includegraphics[width=\linewidth]{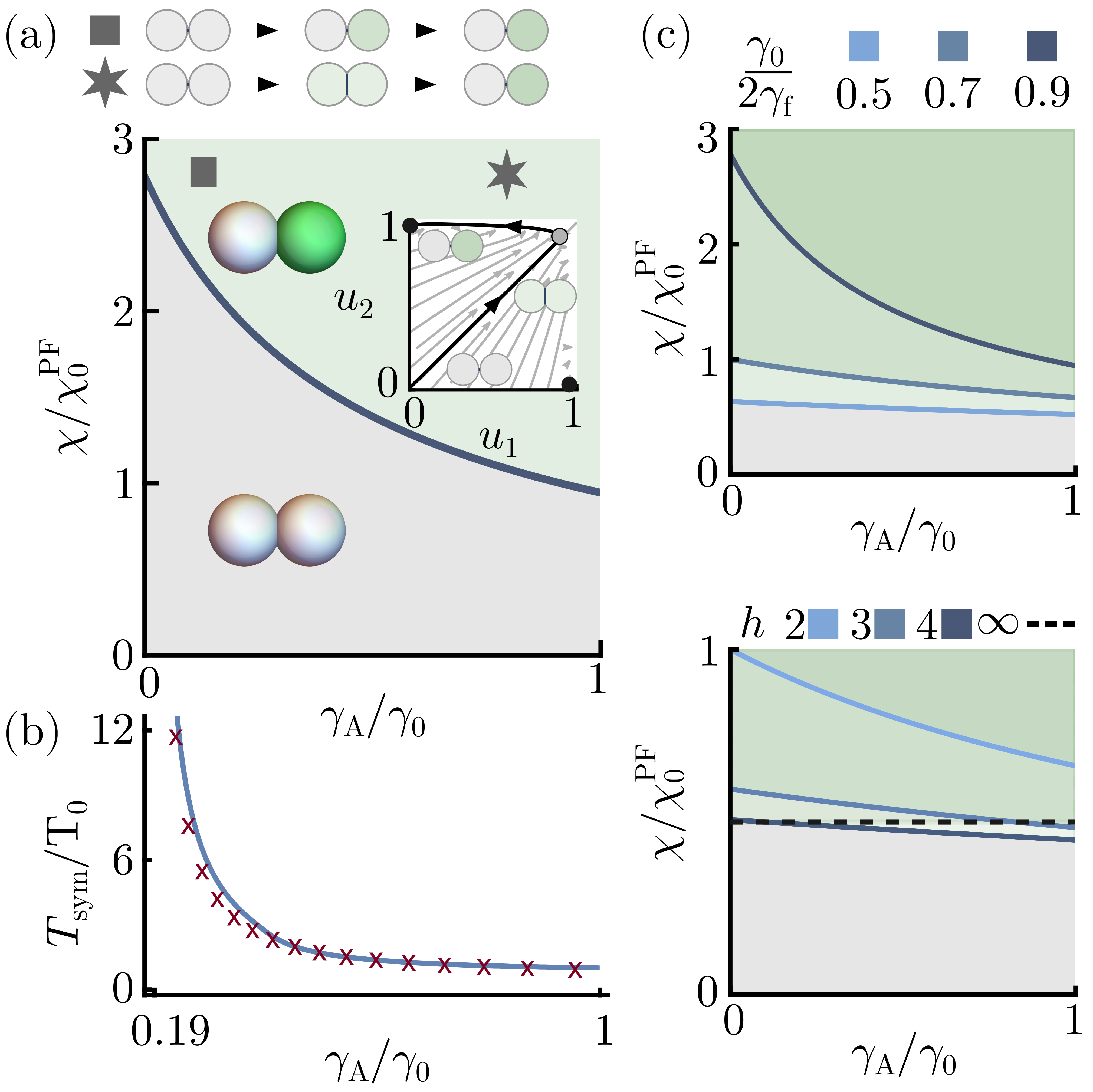}
    \caption{Adaptive adhesion promotes symmetry-breaking. (a) Mutually inhibitory interactions between the signaling states of contacting droplets lead to symmetry-breaking via a line of pitchfork bifurcations (PF), separating uniform (gray) and symmetry broken (green) steady states. Adaptive adhesion promotes symmetry-breaking by transiently increasing the contact area across which mutually inhibitory signals are exchanged (inset: filled black circle: stable steady state, filled gray circle: saddle). (b) The relaxation time to the symmetry-broken steady state $T_{\rm{sym}}$  (blue curve), is dominated by the inverse of the maximum saddle eigenvalue (red crosses) and decreases with increasing $\gamma_{\rm{A}}$, because the larger transient interface allows for the exchange of stronger signals promoting symmetry-breaking ($\chi/\chi_{\rm{0}}^{\rm{PF}}=2$, $T_0$: reference relaxation time for $\gamma_{\rm{A}}/\gamma_0=1$).   
    (c) The baseline tension ratio $\gamma_0/2\gamma_{\rm{f}}$ (top) and the Hill coefficient $h$ of the nonlinear response function [Eq.\eqref{eq:sigma}] (bottom) determine how the critical susceptibility changes with the adaptive adhesion coefficient ($\chi_0^{\rm{PF}}$: reference susceptibility at PF for $\gamma_{\rm{A}}=0$, $\gamma_{\rm{0}}/2\gamma_{\rm{f}}=0.7$, $h=2$).
    Parameter values for each diagram listed in Tab.~\ref{tab:parameter_values_figures}.
    }
    \label{fig:4}
\end{figure}

\begin{figure*}[ht!]
    \centering
    \includegraphics[width=17.8cm]{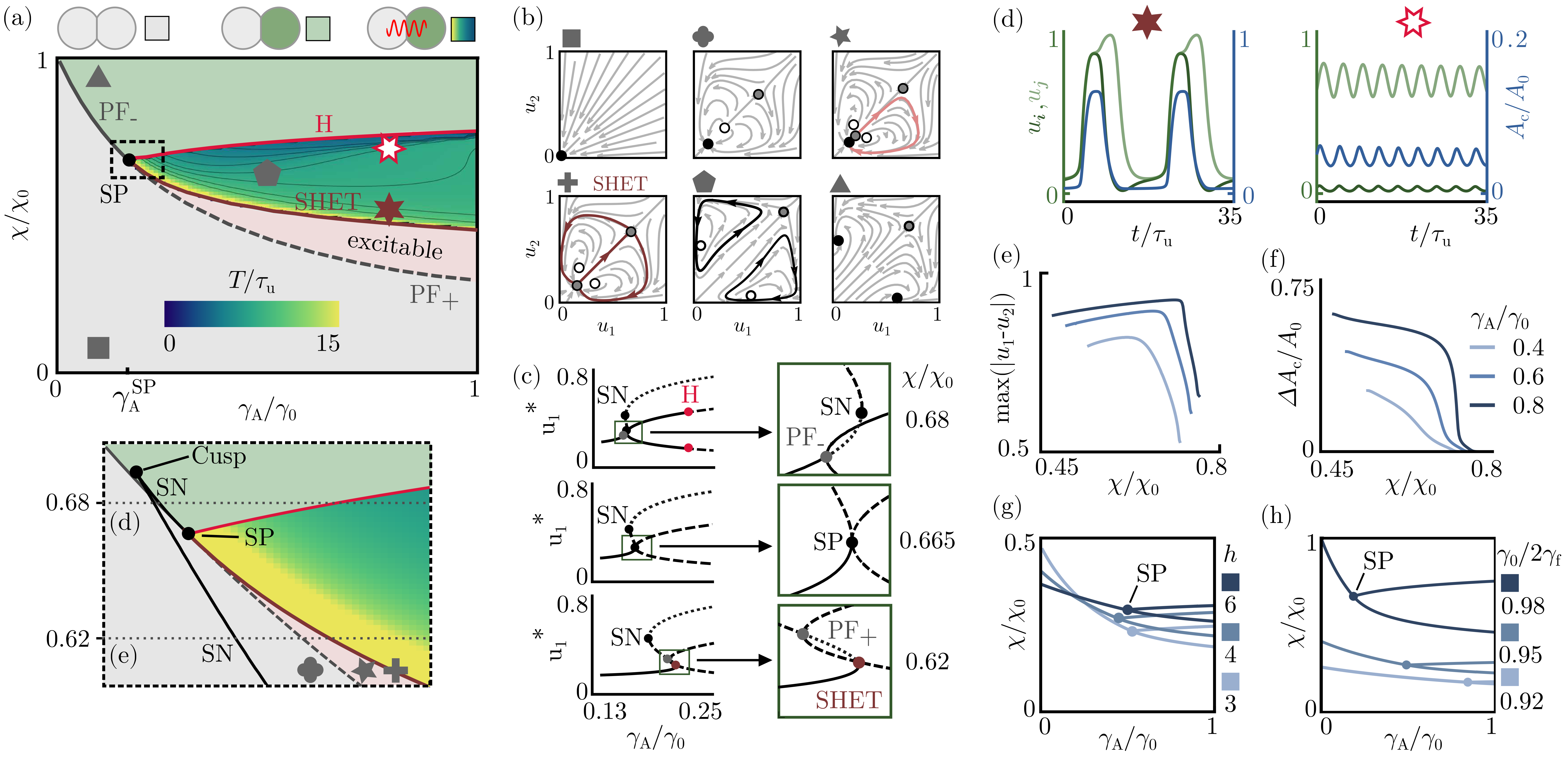}
    \caption{Excitability and oscillations of mechanochemically coupled cell pairs. (a) 
   Adaptive adhesion leads to self-sustained oscillations of signals and cell shapes (color gradient and contour lines denote the oscillation period $T$). The oscillatory regime is surrounded by saddle-heteroclinic (SHET) and Hopf (H) bifurcation lines, which originate from a saddle-node pitchfork codimension-2 point (SP) (PF$_-$: supercritical pitchfork PF$_+$: subcritical pitchfork). Bottom panel: Enlarged view of the SP point environment shows saddle-node (SN) and cusp bifurcations that preserve stable attractor structures. The reference susceptibility is the critical value in the absence of adaptive tension ($\chi_0=\left.\chi_{\rm{PF}}\right|_{\gamma_{\rm{A}}=0}$). (b) Phase portraits for parameter values marked with gray symbols (filled black circles: stable steady states, filled gray circles: saddles, open circles: unstable steady states, rose line: trajectory in the excitable regime, red lines: heteroclinics, black lines: limit cycles). (c) Stable (solid line), unstable (dashed line) fixpoints and saddles (dotted line) computed for variation of $\gamma_{\rm{A}}/\gamma_0$ as indicated by gray dotted lines in (a). Panels on the right show how the PF and SN interact, turning the latter into a SHET (d to e). (d) Oscillation amplitudes decrease and the oscillation period increases with waveforms changing from relaxation-like (near the SHET line) to sinusoidal (near the Hopf line) for increasing $\chi$. (e,f) Maximum difference between internal states (e) and amplitude of contact area changes (f) during oscillations (g,h) The location of the SP point and associated bifurcations in the state diagram depends on the hill coefficient $h$ of the response function (g) [Eq.~\eqref{eq:sigma}] and the baseline tension ratio $\gamma_0/2\gamma_{\rm{f}}$ (h). 
   Parameter values given in Tab.~\ref{tab:parameter_values_figures}. Figure 5 is also shown in the companion letter with minor differences~\cite{PRLJoint} }
    \label{fig:5}
\end{figure*}

The macroscopic equations~\eqref{eq:u_ode},\eqref{eq:signal_definition_singlet},\eqref{eq:signal_definition_doublet},\eqref{eq:adaptive_adhesion_singlet},\eqref{eq:adaptive_adhesion_doublet} and \eqref{eq:contact_area} describe the mechanochemical dynamics of shape-adaptive cells, with two feedback parameters that couple shape changes to signaling (susceptibility $\chi$) and signaling to shape adaptation (adhesion coefficient $\gamma_{\rm{A}}$) [Fig.~\ref{fig:2}(a)], which we have derived from microscopic relations (Secs.~\ref{sec:microscopic_dynamics}-\ref{sec:coarse_grained_feedback}). 
Using a combination of linear stability analysis, simulations, and numerical continuation (Appendix~\ref{sec:numerical_methods} for details), we analyse the dynamical states emerging from this interplay between shape changes and signaling. 
We find that positive feedback between contact-dependent signals and area-increasing adhesion can produce shape bistability, leading to multiple stable wetting states for single droplets on functionalized substrates [Fig.~\ref{fig:3}], and lowering the threshold susceptibility for symmetry-breaking of internal states in interacting pairs of cells [Figs.~\ref{fig:4}]. 
For large adaptive adhesion coefficients moreover, we show that mechanochemical feedback drives excitability and self-sustained oscillations of shapes and internal states [Fig.~\ref{fig:5}]. Finally, we discuss how heterogeneities in mechanical and signaling properties affect these feedback dynamics [Fig.~\ref{fig:6}-\ref{fig:7}].

\subsection{Feedback between contact-based signaling and adaptive mechanics creates bistability}
Equations~\eqref{eq:u_ode},\eqref{eq:signal_definition_singlet},\eqref{eq:adaptive_adhesion_singlet}, and \eqref{eq:contact_area} describe the dynamics of the signaling state $u$ and contact area $A_\mathrm{c}$ of the single, adherent cell.

Depending on the ombination of feedback parameters $\chi$ and $\gamma_\mathrm{A}$ relative to the tension ratio $(\gamma_0-\gamma_{\rm{m}})/\gamma_{\rm{f}}$, Eq.~\eqref{eq:u_ode} has either one or two stable steady state solutions $u^*$ [Fig.\ref{fig:3}(a)]. Using numerical continuation, we find a bistable regime above a critical value of the adaptive adhesion coefficient $\gamma_A^\mathrm{cusp}$ where two saddle-node bifurcation lines (SN) emerge from a cusp bifurcation point [Fig.\ref{fig:3}(a)]. For $\gamma_{\rm{A}}>\gamma_A^\mathrm{cusp}$ and small $\chi$, the only stable solution is a configuration with small contact area $A_{\mathrm{c}}$, correspondingly weak signal transmission and a low signaling state $u$. For values of $\chi$ above the lower SN line, a second stable configuration appears with large contact area $A_{\mathrm{c}}$, which permits a stronger signaling interaction with the substrate and a larger signaling state $u$ [Fig.\ref{fig:3}(a), inset]. This latter configuration is accessible only when the positive feedback between signaling and adaptive mechanics is sufficiently strong. 
The position of the cusp point within the feedback-parameter diagram, and the size of the associated bistable regime depends on the tension ratio $(\gamma_0-\gamma_{\rm{m}})/\gamma_{\rm{f}}$, and on the Hill coefficient $h$ in the response function [Eq.~\eqref{eq:sigma}]---increasing either of the two parameters lowers the threshold adaptive adhesion coefficient $\gamma_A^\mathrm{cusp}$ [Fig.~\ref{fig:3}(b)]. We find bistability for $h\geq2$.

In the limit $h\rightarrow\infty$, i.e. where the internal states respond to signals in a step-wise manner [Fig.~\ref{fig:2}(b)], one can derive a simple relation between $\chi$ and $\gamma_A$ for the two saddle-node lines [Fig.~\ref{fig:3}(c)]. In this limit, the only possible stable steady state solutions of Eq.~\eqref{eq:u_ode} are $u^*\in\{0;1\}$ and the corresponding surface tensions at the contact site are $\gamma_c \in \{\gamma_0;\gamma_0 - \gamma_A\}$ [Eq.~\eqref{eq:adaptive_adhesion_singlet}]. For small values of $\chi$, signaling is weak and the only stable steady state is $u^*=0$ with a small contact area set by $\gamma_c = \gamma_0$. The second stable steady state $u^*=1$ appears for
\begin{align}
    s(\left. A_\mathrm{c}\right|_{\gamma_{\rm{c}}=\gamma_0-\gamma_A}) \geq 1.
    \label{eq:SN2_condition}
\end{align}
For
\begin{align}
    s(\left. A_\mathrm{c}\right|_{\gamma_{\rm{c}}=\gamma_0}) > 1,
    \label{eq:SN1_condition}
\end{align}
the configuration with small contact area and $u=0$ is no longer a steady state solution and $u^*=1$ remains the only stable steady state. 
From conditions ~\eqref{eq:SN2_condition}--\eqref{eq:SN1_condition} together with \eqref{eq:signal_definition_singlet} follows that the critical susceptibilities at the saddle-node lines delineating the bistable regime are given by 
\begin{align}
    \chi_1 = \frac{
   A_0
    }{
    \left. A_{\mathrm{c}}\right|_{\gamma_\mathrm{c} = \gamma_0-\gamma_{\mathrm{A}}}
    }
    \label{eq:single_droplet_sn1}
\end{align}
and
\begin{align}
    \chi_2 = \frac{
   A_0
    }{
    \left. A_{\mathrm{c}}\right|_{\gamma_\mathrm{c} = \gamma_0}
    }.
    \label{eq:single_droplet_sn2}
\end{align}
for the lower and upper lines respectively [Fig.~\ref{fig:3}(c)]. 

\subsection{Symmetry-breaking of internal states}

Next, we study the dynamics of cell pairs exchanging mutually inhibitory signals Eqs.~\eqref{eq:u_ode},\eqref{eq:signal_definition_doublet}, \eqref{eq:adaptive_adhesion_doublet}, and \eqref{eq:contact_area}. Strong mutually inhibitory interactions generically lead to spontaneous symmetry-breaking, whereby initially small differences between interacting units diverge to low- and high-value steady states \cite{pisarchik2022multistability}, a mechanism relevant for the patterning of different cell types \cite{sjoqvist2019say}. Using numerical continuation, we find that in the state-diagram of feedback parameters the regimes of uniform and symmetry-broken steady states are separated by a line of supercritical pitchfork bifurcations (PF) [Fig.~\ref{fig:4}(a)]. Below the critical value $\chi_{\rm{PF}}$, inhibition is not strong enough to produce symmetry-breaking, and the cell pair converges to identical low internal states with a small contact area. Linear stability analysis shows that this critical susceptibility scales approximately inversely with the interfacial area $\chi_{\rm{PF}} \sim A_0/A_c$ (Appendix~\ref{sec:approximation_PF})---indicating that the adaptive adhesion promotes symmetry-breaking: starting from low, nearly identical internal states, the active term in Eq.~\eqref{eq:contact_tension_doublet} transiently expands the contact area as the trajectory approaches a saddle in the phase space of internal states [Fig.~\ref{fig:4}(a)(inset)]. The large contact effectively lowers the threshold susceptibility and drives the divergence of the internal states, which in turn reduces adhesion and the contact area. Correspondingly, starting from nearly uniform conditions, the time it takes for the internal states to diverge decreases with increasing $\gamma_{\rm{A}}$ and correlates with the largest eigenvalue of the saddle [Fig.\ref{fig:4}(b)]. We find regimes of symmetry-breaking for $h\geq2$, which increase with the baseline tension ratio $\gamma_0/2\gamma_{\rm{f}}$ [Fig.~\ref{fig:4}(c)], as well as with increasing Hill coefficient $h$ [Fig.~\ref{fig:4}(d)]. 
Overall, we find that shape-dependent mechanochemical feedback increases the robustness of symmetry-breaking, which could aid reliable fate determination in noisy biological environments \cite{dullweber2023mechanochemical, erzberger2020mechanochemical, cohen2023precise}. For instance, adaptive contact dynamics occur between sensory cell pairs in zebrafish embryos that exchange mutually inhibitory signals to undergo robust symmetry breaking~\cite{jacobo2019notch, erzberger2020mechanochemical}.

\subsection{Tunable self-sustained oscillations}
\label{sec:oscillations_symmetric}
At large values of the adaptive adhesion coefficient, the coupling between signaling and interface geometry can drive self-sustained oscillations of the internal states and shape of interacting cell pairs.
These oscillations are driven by competition between the adaptive adhesion and the tendency of the pair to undergo symmetry-breaking: the product of internal states $u_1u_2$ increases the contact area according to Eq.\eqref{eq:contact_tension_doublet}, thereby driving their own inhibition, leading to negative feedback.
The oscillatory regime, bounded by Hopf (H) and saddle heteroclinic (SHET) bifurcation lines, separates the stable symmetric and symmetry-broken states in the parameter diagram [Fig.~\ref{fig:5}(a-b)], derived via numerical continuation (Appendix~\ref{sec:numerical_methods}). 
These lines originate from a saddle-node pitchfork bifurcation point (SP)---a codimension-2 bifurcation at which the PF line tangentially intersects with a saddle-node (SN) bifurcation line [Fig.~\ref{fig:5}(a,c) and Fig.~\ref{fig:SPDetails}] \cite{blackbeard2012bursting}. 

When $\gamma_{\rm{A}}>\gamma_{\rm{A}}^{\rm{SP}}$ and $\chi$ reaches the critical susceptibility $\chi_{\rm{PF}}$, the inhibitory signals induce symmetry-breaking and the unstable fixed point undergoes a subcritical pitchfork bifurcation, producing a saddle and two new unstable fixed points [Fig.~\ref{fig:5}(b) star]. In this regime the droplet pair is excitable: fluctuations moving the internal states beyond the separatrices, which connect the saddle to the unstable fixed points, trigger a large increase of both internal states and the contact area $A_{\rm{c}}$, followed by transient symmetry-breaking [Fig.~\ref{fig:5}(b) star and Movie~1 (b)]. Increasing $\chi$ shortens the distance between the uniform stable fixed point and the saddle, thus lowering the excitation threshold until the two points collide at the SHET line and give rise to a pair of heteroclinic orbits that connect the resulting transversely stable, nonhyperbolic point to the second saddle point [Fig.~\ref{fig:5}(b) cross]. This nonhyperbolic point is destroyed as the heteroclinic orbits bifurcate into two symmetric stable limit cycles [Fig.~\ref{fig:5}(b) pentagon and Movie~1 (f)], which remain the only stable attractors of the system. Thus, cycles appear once transmitted signals are strong enough to induce symmetry-breaking, which in turn lowers the adhesion---and thereby the contact area---sufficiently to reduce signals below the symmetry-breaking threshold. In turn, the product of states [Eq.~\eqref{eq:adaptive_adhesion_doublet}] increases again, thereby driving adhesion, contact area, and signal amplitude back above the threshold. ing on the two feedback parameters, the mechanochemical oscillations exhibit a range of temporal profiles. Near the SHET line the droplet pair exhibits relaxation-type oscillations in which it spends a large fraction of the cycle in small-area configurations with nearly identical states, interrupted by spikes in the contact area $A_{\rm{c}}$ and rapid, transient symmetry-breaking [Fig.~\ref{fig:5}(d) and Movie~1 (c)]. The oscillation period diverges as $\chi$ approaches $\chi_{\rm{SHET}}$ due to the ghost of the destroyed saddle point that critically slows down the limit-cycle phase when passing through its vicinity [Fig.~\ref{fig:5}(a)]. With increasing $\chi$, the time-averaged difference between the internal states increases and the oscillation amplitudes decrease, reaching near-sinusoidal waveforms in states and contact area close to the Hopf bifurcation line [Movie~1 (d)], where the limit cycles smoothly contract into symmetry-broken fixed points [Fig.~\ref{fig:5}(b,d-e) and Movie~1 (f)].

The position of the SP point within the feedback-parameter diagram, and the size of the associated regimes depend on the baseline tension ratio $\gamma_0/2\gamma_{\rm{f}}$, and on the Hill coefficient $h$ in the response function [Eq.~\eqref{eq:sigma}]. Increasing $\gamma_0$ lowers the threshold adaptive tension for the onset of oscillations [Fig.~\ref{fig:5}(h)], while for low $\gamma_0$ the adaptive adhesion can push the interface into a regime where any area increase lowers the total surface energy, leading to shape instabilities~\cite{bormashenko2017physics, binysh2022active}. Close to $\gamma_{\rm{A}}/\gamma_0=1$, such instabilities may remain transient, i.e. restricted to fractions of the oscillation phase, before restabilizing due to the decrease of adhesion upon symmetry-breaking of internal states, whereas at large $\gamma_{\rm{A}}/\gamma_0$, these effects are expected to dominate the dynamics and lead to new phenomena.

We found shape bistabilities and symmetry-breaking for Hill coefficients $h\geq2$, and oscillations for $h\geq3$. Strongly nonlinear response functions are commonly used to model regulatory feedbacks in cells~\cite{boareto2015jagged, shi2018understanding, corson2017self}, and experimental evidence has been reported for e.g. the Nodal pathways \cite{dubrulle2015response,PRLJoint}. Interestingly, we observe that strong adaptive adhesion achieves lower thresholds for smaller Hill coefficients, i.e. that the PF bifurcation lines for different Hill coefficients intersect in the feedback parameter space [Fig.~\ref{fig:4}(c), Fig.~\ref{fig:5}(g)], indicating a non-trivial interplay between the response nonlinearity and the geometry-dependent nonlinearity which together drive symmetry-breaking.

Together, these results illustrate how mechanochemical feedback can drive excitability and self-sustained oscillations.

\begin{figure}[ht!]
    \centering
    \includegraphics[width=\linewidth]{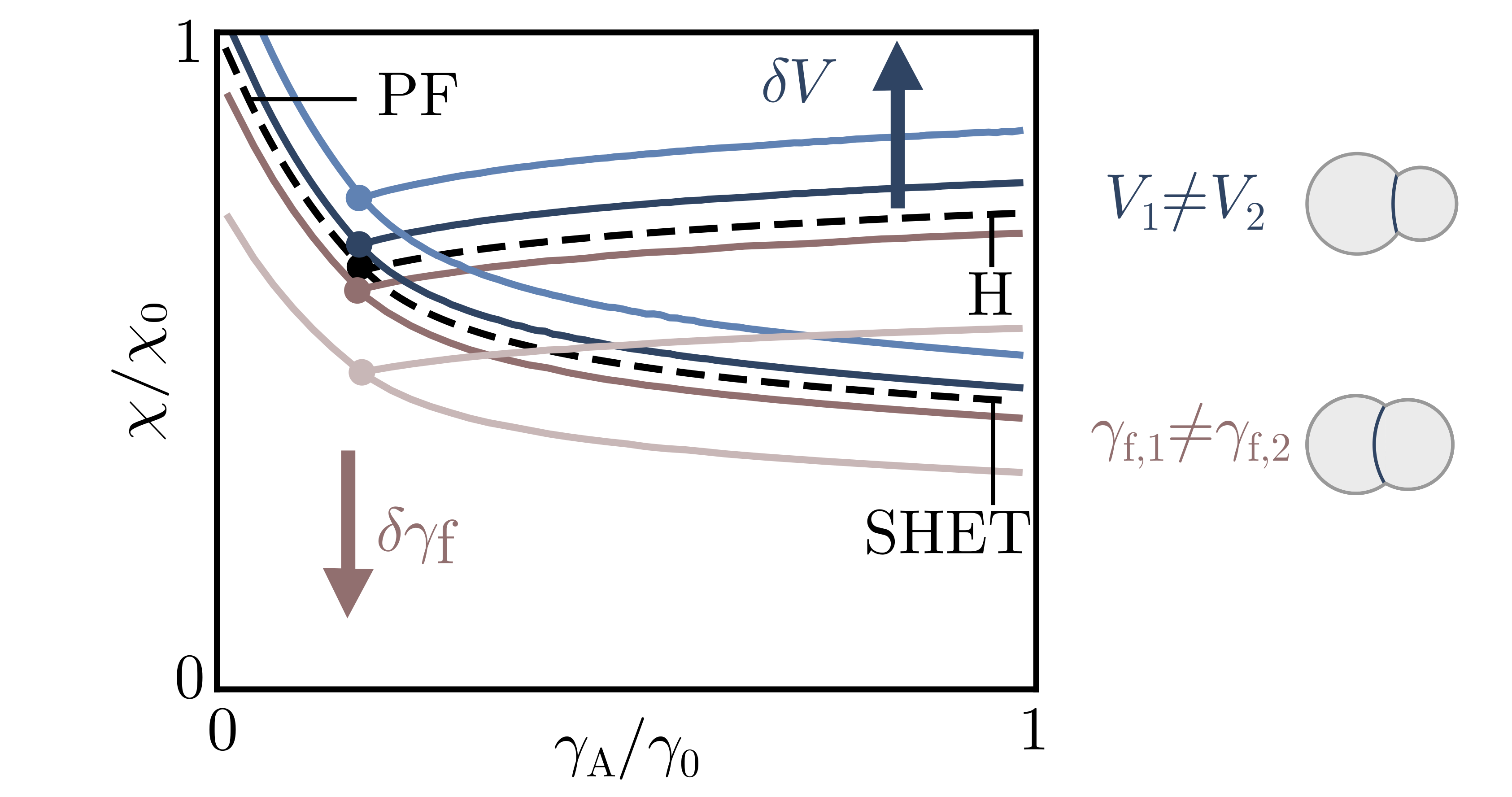}
    \caption{Different cell volumes $\delta V/\bar{V}=\{0.25,0.5\}$ (blue) or outer tensions $\delta \gamma_{\rm{f}}/\bar{\gamma_{\rm{f}}}=\{0.25,0.5\}$ (brown) shift the SP point and associated bifurcation lines in the state diagram. Parameter values given in Tab.~\ref{tab:parameter_values_figures}.}
    \label{fig:6}
\end{figure}

\begin{figure}[ht!]
    \centering
    \includegraphics[width=\linewidth]{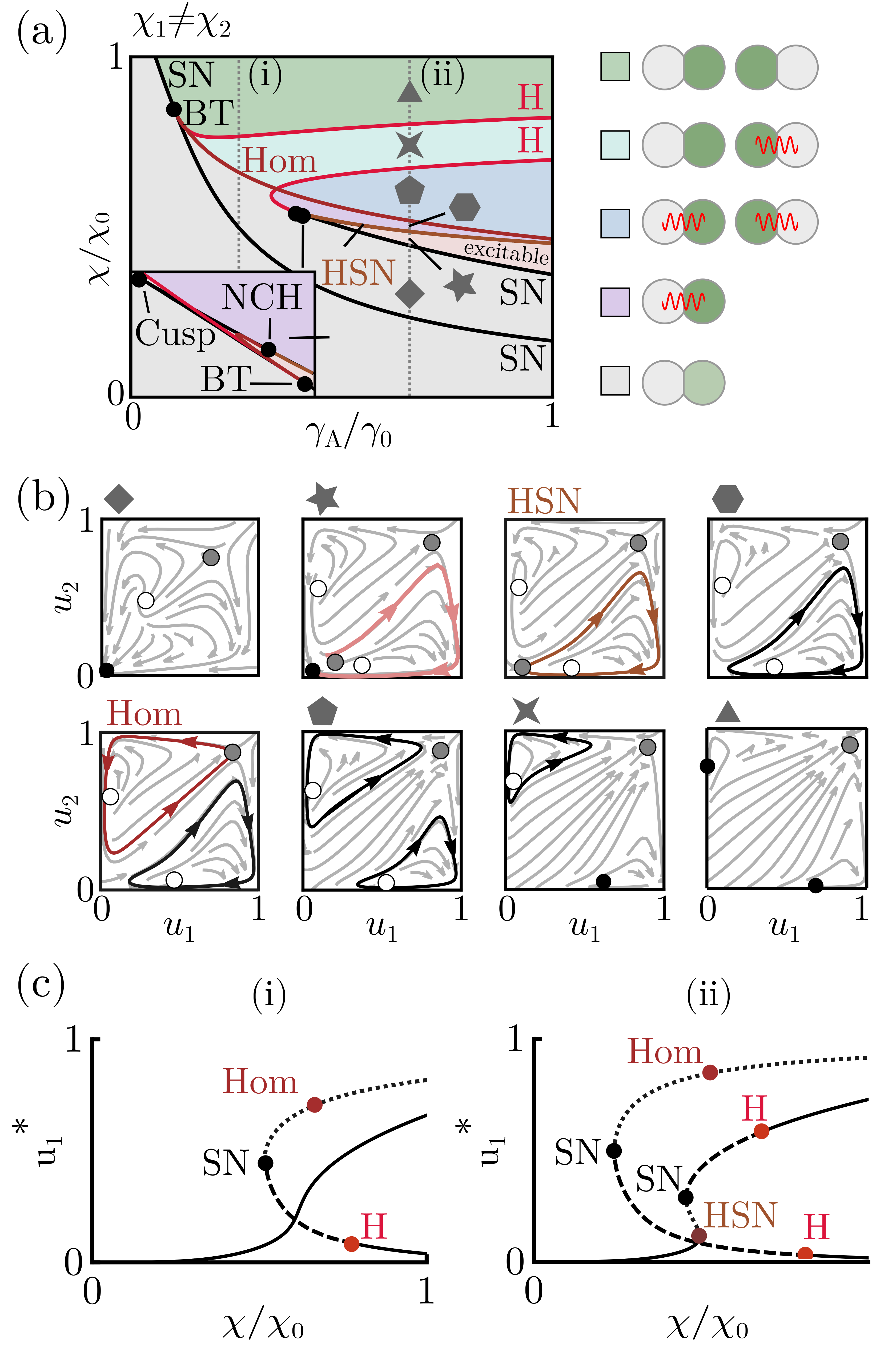}
    \caption{Heterogeneous susceptibility in interacting cell pairs. (a) A difference in signal susceptibilities $\delta \chi/\bar{\chi}=0.05$ (i.e. $\chi_1/\chi_2 \approx 1.1$) splits the SP point into a pair of Bogdanov-Takens bifurcation points (BT), a non-central homoclinic to saddle-node bifurcation (NCH) and associated bifurcation lines. Inset shows the state diagram close to the second BT point ($\gamma_{\rm{A}}/\gamma_0\in[0.3896,0.403]$, $\chi/\chi_0\in[0.5388,0.5441]$). Note that NCH and BT are connected by a homoclinic (Hom). (HSN: Saddle-node homoclinic). (b) Phase portraits for parameter values marked with gray symbols in (b). (filled black circle: stable steady state, filled gray circle: saddle, open circle: unstable steady state, rose line: trajectory in the excitable regime, thick black line: limit cycle) (c) Wit unequal signaling properties ($\chi_1 \neq \chi_2$), the pitchfork bifurcation is replaced by a new saddle-node bifurcation (compare to Fig.~\ref{fig:5}(c)).
    Parameter values given in Tab.~\ref{tab:parameter_values_figures}.}
    \label{fig:7}
\end{figure}

\subsection{Droplet heterogeneities}
The SP point arises for identical droplets. While such state-space structures have been found and experimentally characterized for instance in optical cavities \cite{blackbeard2012bursting}, most physical systems exhibit non-negligible variations in their properties. Differences in the properties of the interacting droplets change the state diagram shown in Fig.~\ref{fig:5}(a). For unequal droplet volumes $V_{1,2} = \bar{V} \pm \delta V$, symmetry-breaking and oscillatory dynamics emerge at a larger signaling susceptibility $\chi$ than in pairs of identical droplets, whereas a difference in the outer surface tensions $\gamma_{\rm{f},1,2} = \bar{\gamma_{\rm{f}}} \pm \delta \gamma_{\rm{f}}$ promotes symmetry-breaking and oscillations at lower susceptibilities due to partial internalization resulting in larger equilibrium contact areas [Fig.~\ref{fig:6}]. 

Tension and volume asymmetry do not favour any droplet to reach a higher or lower internal state, because the signaling properties of each droplet remain unaffected, and thus the topology of the state space is preserved. In contrast, numerical continuation shows that a difference in the signaling susceptibility $\chi_{1,2} =  \bar{\chi} \pm \delta \chi$ splits the SP point into two Bogdanov-Takens (BT) codimension-2 points, and the SHET line into two homoclinics (Hom) and a saddle-node homoclinic (HSN) bifurcation line emerging from a non-central homoclinic to saddle-node (NCH) [Fig.~\ref{fig:7}(a)]. 
Accordingly, the limit cycle and the corresponding symmetry-broken state, in which the less susceptible droplet maintains the lower internal state, require lower values of $\chi$ and $\gamma_{\rm{A}}$ than the inverse symmetry-broken states. Thus, two limit cycles appear at different susceptibilities through a HSN and a Hom bifurcation [Fig.~\ref{fig:7}(b)], compared to homogeneous droplets, for which two limit cycles appear simultaneously in a SHET bifurcation [Fig.~\ref{fig:5}(b) cross]. This allows for parameter regimes with single limit cycles [Fig.~\ref{fig:7}(b) hexagon] or coexistence with stable fixed points [Fig.~\ref{fig:7}(b) 4-pointed star]---contrary to the case of identical susceptibilities. Heterogeneous material properties can thus produce an even wider spectrum of dynamics.

\section{Discussion}

Motivated by mechanochemical feedback in biological cells, we derive a tractable set of macroscopic equations from underlying microscopic dynamics of contact-dependent biochemical signaling and adhesion [Figs.~\ref{fig:1}-\ref{fig:2}]. Specifically, we consider cell-cell signals exchanged at contact interfaces, whose magnitude depends on the contact area, and which drive changes in cell-cell adhesion, thereby feeding back onto the exchanged signals.
We focus on systems in which the signals trigger an internal response that evolves slowly compared to the cell shape dynamics, i.e. controlled by the transcriptional regulation of production and decay processes. 

We find that the self-amplification of area-dependent signals, which increase adhesion at the signaling interface leads to shape bistability once the system crosses a saddle-node bifurcation line, set by the strength of adhesion adaptation and signal susceptibility
[Fig.~\ref{fig:3}]. Moreover, in the presence of mutually inhibitory signaling interactions, shape-dependent feedback gives rise to robust symmetry-breaking [Fig.~\ref{fig:4}], excitability, and self-sustained oscillations ranging from sinusoidal to relaxation-like waveforms [Fig.~\ref{fig:5}]. 
While mutual inhibition can lead to symmetry-breaking of internal states even in the absence of shape-dependent feedback, the adaptive tension lowers the critical susceptibility at which the internal states diverge, i.e. where the system crosses a line of pitchfork bifurcations [Fig.~\ref{fig:3}]. Furthermore, because mutual inhibition lowers adhesion, but adhesion in turn increases contact area-dependent signals, the competition between signal self-amplification and symmetry-breaking can produce oscillatory dynamics for strong adaptive adhesion. In particular, these dynamics are possible once the adaptive adhesion coefficient exceeds a threshold value at which the saddle-node bifurcation line associated with the shape bistability collides with the pitchfork bifurcation line associated with symmetry-breaking, giving rise to a new codimension-2 point [Fig.\ref{fig:5}(a,c)].

Oscillatory signaling dynamics arise in diverse multicellular systems \cite{goldbeter1997biochemical}, and are important for the determination of cell fates during embryonic development or stem cell maintenance \cite{lewis2003autoinhibition, ferrell2011modeling, nandagopal2018dynamic, simon2020live, raina2022intermittent}. While these dynamics are primarily thought of as purely biochemical phenomena \cite{casani2022signaling,schreiber2019reaction}, our work reveals that the coupling to cellular shapes can produce such complex behaviors with a minimal set of degrees of freedom and macroscopic control parameters. 
It will be interesting to investigate if mechanochemical oscillations are more robust to perturbations and noise compared to purely biochemical oscillations. Moreover, mechanochemical coupling allows a self-organizing system to coordinate state transitions (e.g. cell fate decisions) with spatial rearrangements in noisy environments without requiring upstream control mechanisms. For example, by creating and removing specific contact surfaces, cells can robustly undergo a sequence of distinct differentiation events \cite{erzberger2020mechanochemical, priya2020tension, cohen2023precise, dray2021dynamic}.%

Proximity to critical points allows biological systems to undergo abrupt transitions in their macroscopic properties in response to small microscopic changes, allowing for very sensitive adaptation \cite{petridou2021rigidity}. For example, self-amplifying mechanochemical coupling allows cells in developing zebrafish embryos to undergo a switch-like change from low-adhesion to high-adhesion configurations in distinct tissue regions, facilitating the formation of fate boundaries~\cite{PRLJoint,petridou2021rigidity}. Similarly, synthetic tissues \cite{toda2018programming}, biomimetic droplets \cite{gonzales2023bidirectional}, and materials that exhibit adaptive shape responses could be engineered to harness geometry-dependent feedback loops to achieve self-assembling and self-healing functionalities \cite{mcevoy2015materials, luo2023autonomous,shah2021shape, sun2023mean, terryn2021review}.

Our work focuses on signaling dynamics which trigger a slow response, set e.g. by the timescale of transcriptional regulation, compared to the frictional timescales determining the shape dynamics. 
Faster responses are possible when signaling interactions drive local biochemical changes (e.g. through the phosphorylation of stress regulators in the cytoskeleton \cite{garrido2021nonmuscle}). In such systems, the shape becomes a dynamic degree of freedom, capable of acting as a form of memory, and supporting further dynamical regimes controlled by the combination of frictional and biochemical timescales. 

In conclusion, by bridging the microscopic and macroscopic scales, our model provides a framework for understanding mechanochemical feedback in soft active materials, revealing universal characteristics that could have significant implications for both biological and synthetic systems. 

\paragraph*{Acknowledgements}
We gratefully acknowledge insightful discussions and valuable feedback from Florian Berger, Erwin Frey, Isabella Graf, Jeremy Gunawardena, Adrian Jacobo, Thomas Quail, Ulrich Schwarz, Alejandro Torres-Sánchez, Falko Ziebert and all members of the Erzberger Group. We acknowledge funding by the EMBL and TD was supported by a Joachim-Herz Add-on fellowship for interdisciplinary life science. We thank Alba Diz-Mu\~noz for supporting the acquisition of microscopy image Fig.~\ref{fig:1}(a).


\appendix
\section{Statistical physics of adhesion molecule binding}
\label{sec:adhesion_statmech}

At steady state, the flux coupling bulk and surface concentrations [Eq.~\eqref{eq:bc_flux}] vanishes and the surface can be considered to be in chemical and thermal equilibrium with a constant temperature $T$ and in contact with a bath of constant chemical potential $\mu=\mu(c_\mathrm{N})$ set by the steady state bulk concentration. Note that the chemical potential is kept constant through a non-equilibrium process---the turnover of adhesion molecules. Each binding site at the interface is a two-state system: a binding site is either occupied or unoccupied. If $n$ is the number of occupied binding sites, $n^\mathrm{max}$ the total number of available binding sites at the surface and $\epsilon$ the binding energy, then the grand canonical partition sum for the whole surface reads
\begin{align}
    \Xi = \sum_{n=0}^{n^\mathrm{max}} \binom{n^\mathrm{max}}{n}e^{\beta n(\mu-\epsilon)} = \left( 1+e^{\beta(\mu-\epsilon)} \right)^{n^\mathrm{max}}
\end{align}
with $\beta=(k_{\mathrm{B}} T)^{-1}$. The ensemble average of the number of occupied binding sites is
\begin{align}
    \langle n \rangle = \frac{1}{\beta} \frac{\partial \ln \Xi}{\partial\mu} = \frac{n^\mathrm{max}}{1+e^{\beta(\epsilon - \mu)}},
    \label{eq:fermi_dirac}
\end{align}
showing that the system follows Fermi-Dirac statistics. In the chemical equilibrium, the rates of binding and unbinding must be equal for each binding site. The binding rate of adhesion molecules is
\begin{align}
    k_\mathrm{binding} = k_\mathrm{on}^\mathrm{N} c_\mathrm{N} p_\mathrm{uoc}
    \label{eq:k_binding}
\end{align}
with $p_\mathrm{uoc}=1/(1+e^{\beta(\mu - \epsilon)})$ the probability that a binding site is not occupied, while the unbinding rate is
\begin{align}
    k_\mathrm{unbinding} = k_\mathrm{off}^\mathrm{N} p_\mathrm{oc}
    \label{eq:k_unbinding}
\end{align}
with $p_\mathrm{oc} = e^{\beta(\mu - \epsilon)}/(1+e^{\beta(\mu - \epsilon)})$ the probability that a binding site is occupied. From $k_\mathrm{binding} = k_\mathrm{unbinding}$ and Eqs.~\eqref{eq:k_binding}--\eqref{eq:k_unbinding} follows
\begin{align}
    \frac{k_{\mathrm{off}}^{\mathrm{N}}}{k_{\mathrm{on}}^{\mathrm{N}} c_{\mathrm{N}}} = e^{\beta(\epsilon - \mu)}.
\end{align}
From $m_\mathrm{N} = \langle n \rangle/A_\mathrm{c}$ and $m_\mathrm{N}^\mathrm{max} = n^\mathrm{max}/A_\mathrm{c}$ together with Eq.~\eqref{eq:fermi_dirac} and $c_{\rm{N}}=k_{\rm{p}}^{\rm{N}}/k_{\rm{d}}^{\rm{N}}$ follows then Eq.~\eqref{eq:mN}.

\section{Symmetry-breaking of signaling states}
\label{sec:approximation_PF}

The following Appendix is also included as supplementary material in~\cite{PRLJoint}.

\begin{figure}
    \centering
    \includegraphics{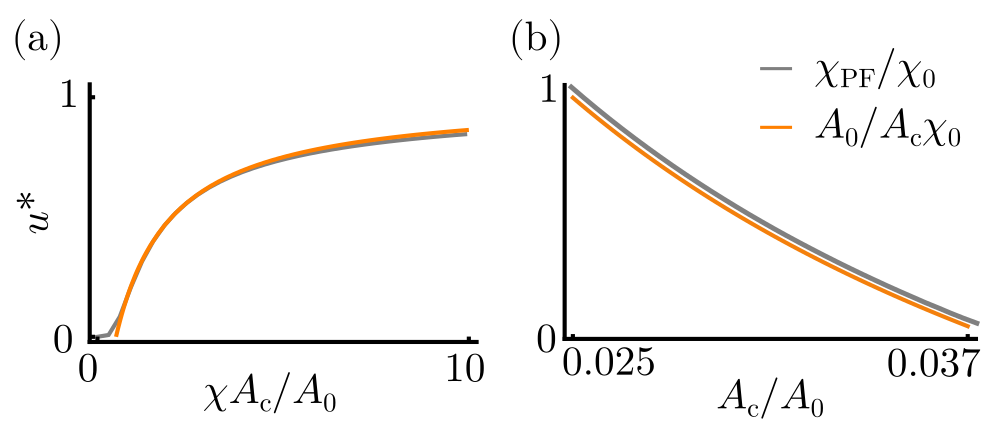}
    \caption{(a) Uniform fixpoints of Eq.~\eqref{eq:u_ode} computed numerically (gray) and approximation from linearization of the response function $\sigma(s_{ij})$ around $s_{ij}=1$ [Eq.~\eqref{eq:uniform_steady_states}], (orange), $h=4$. (b) Comparison between Eq.~\eqref{eq:symmetry_breaking_scaling} (orange) and the steady state contact area computed numerically along the supercritical pitchfork bifurcation line derived via continuation in MatCont (gray). This result is also presented in the supplementary material of the companion letter~\cite[Fig.~7]{PRLJoint}}
    \label{fig:symmetry_breaking}
\end{figure}

\begin{figure}[ht!]
    \centering
    \includegraphics[]{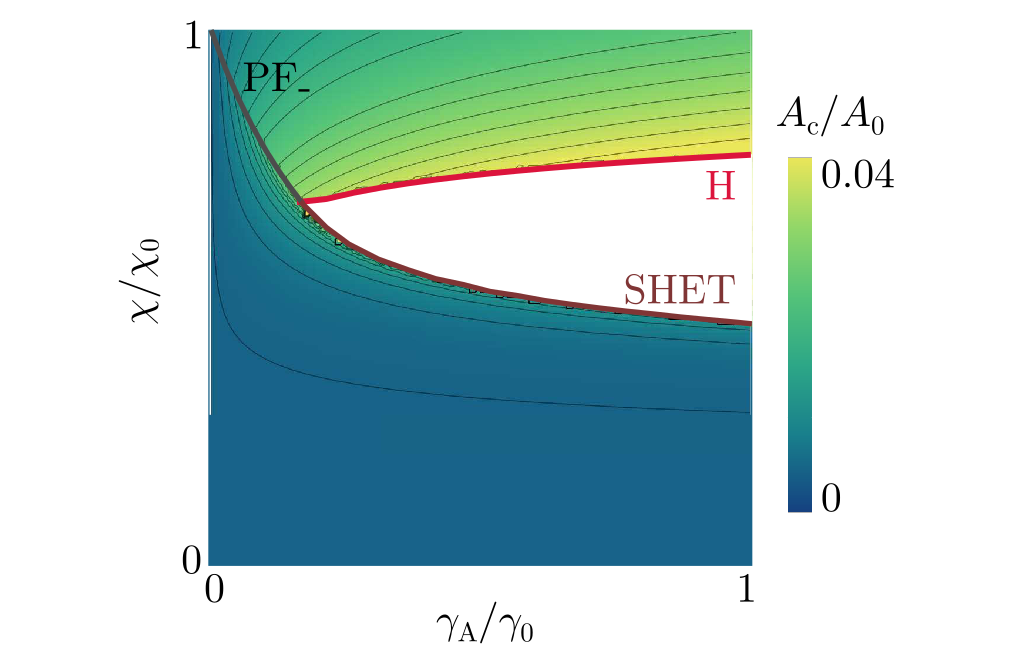}
    \caption{State diagram as shown in Fig.~\ref{fig:5}(a) for $\gamma_0/2\gamma_{\rm{f}} = 0.98$. The color code indicates the normalized contact area at the fixpoint, i.e. at the uniform fixpoint state below the PF and SHET line and at the symmetry-broken fixpoints above the PF and Hopf line. The oscillatory regime between Hopf and SHET line is white as it does not contain any stable fixpoints. This result is also presented in the supplementary material of the companion letter~\cite[Fig.~8]{PRLJoint}.}
    \label{fig:SteadyStateContactArea}
\end{figure}

In many biological systems, Notch signals are mutually inhibitory, i.e. signals suppress the production of ligands \cite{sjoqvist2019say}. Strong mutual inhibitory interactions generically lead to spontaneous symmetry-breaking \cite{pisarchik2022multistability}, whereby small initial differences in the signaling states are amplified and diverge to high- and low-value steady states. At the onset of symmetry-breaking, the uniform steady state solution of Eq.~\ref{eq:u_ode} becomes unstable. To derive an approximation for the onset of symmetry-breaking, we expand $\sigma(s_{ij})$ [Eq.~\eqref{eq:sigma}] for a general Hill coefficient $h$ to first order around the inflection point $s_{ij}=1$ 
\begin{align}
    \sigma(s_{ij}) = \frac{1}{2} + \frac{h}{4}(s_{ij} - 1) + \mathcal{O}((s_{ij}-1)^2)
\end{align}
yielding the dynamic equation
\begin{align}
    \tau_{\rm{u}} \frac{du_i}{dt} = \frac{1}{2} + \frac{h}{4}(s_{ji} - 1) - u_i
    \label{eq:u_ode_aprrox}
\end{align}
and using the definition of the signal Eq.~\eqref{eq:signal_definition_doublet} the uniform steady state is [Fig.~\ref{fig:symmetry_breaking}(a)]
\begin{align}
    u^* = 1 - \frac{2+h}{4 + h \chi \dfrac{A_{\rm{c}}}{A_0}}.
    \label{eq:uniform_steady_states}
\end{align}
Linear stability analysis reveals that this uniform steady state looses stability for
\begin{align}
    \chi_{\mathrm{PF}} = \frac{4 A_0}{h A_{\mathrm{c}}},
    \label{eq:symmetry_breaking_scaling}
\end{align}
with $A_{\mathrm{c}}=A_{\mathrm{c}}(\gamma_{\mathrm{c}})$ and $\gamma_{\mathrm{c}} = \gamma_0 - \gamma_{\mathrm{A}}(u^*)^2$. 
Comparison with the steady state contact area computed numerically along the supercritical pitchfork bifurcation line that was derived via continuation in MatCont shows good agreement [Fig.~\ref{fig:symmetry_breaking}(b)]. Figure~\ref{fig:SteadyStateContactArea} shows the normalized steady state contact area $A_{\rm{c}}/A_0$ in the state space of feedback parameters.

\section{Shapes of asymmetric droplets}
\label{sec:shape_asymmetric_droplets}

\begin{figure*}
    \centering
    \includegraphics{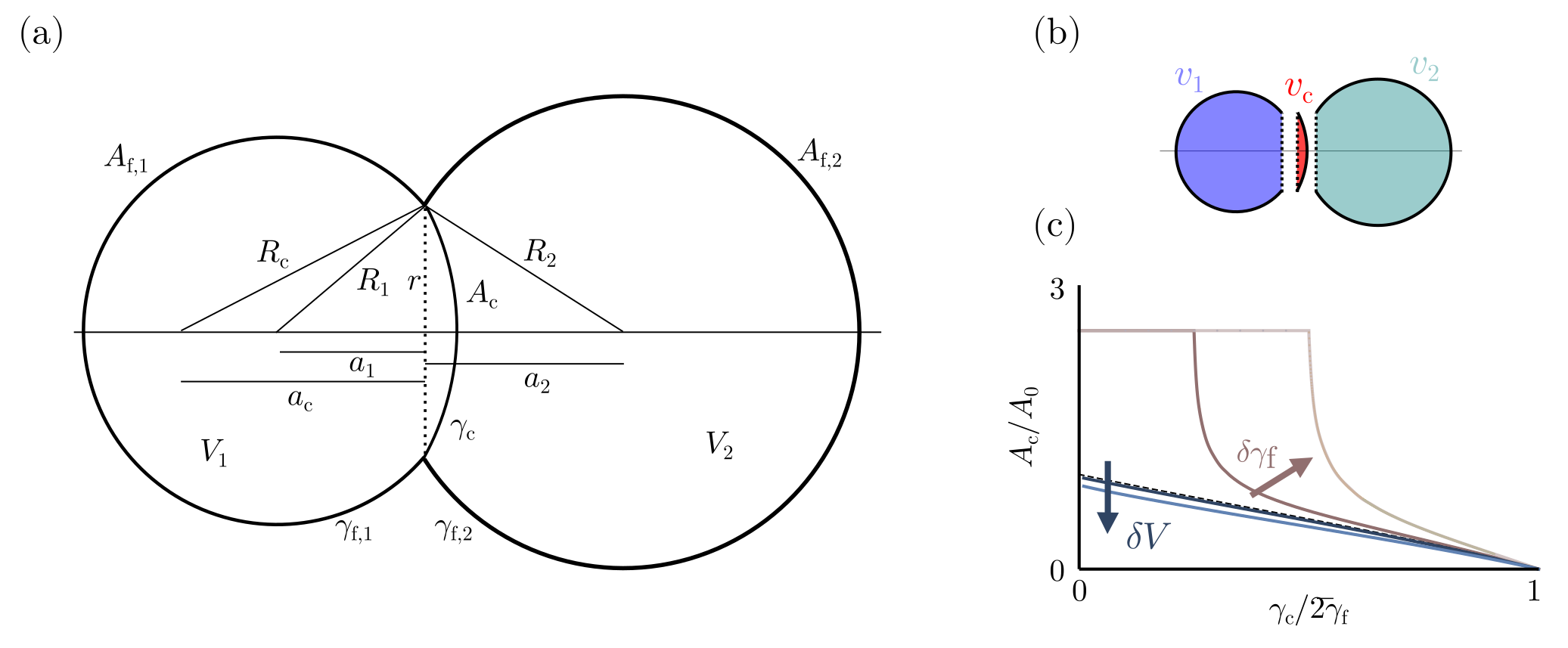}
    \caption{ (a) Parameterization of a pair of asymmetric droplets adapted from \cite{maitre2016asymmetric}. (b) The droplet volumes can be expressed in terms of the three spherical cap volumes  $v_1,v_2,v_{\rm{c}}$ (c) Differences in droplet volumes $\delta V/\bar{V}=\{0.25,0.5\}$ (blue) or outer interfacial tensions $\delta \gamma_{\rm{f}}/\bar{\gamma_{\rm{f}}}=\{0.25,0.5\}$ (brown) change how the contact area between the droplets depends on the tension ratio $\gamma_{\rm{c}}/2\bar{\gamma}_{\rm{f}}$. Curves are obtained by numerically minimizing Eq.~\eqref{eq:energy_asymmetric_doublet_2} (Appendix~\ref{sec:shape_asymmetric_droplets} for details) }
    \label{fig:asymmetry_doublet_parameterization}
\end{figure*}

For pairs of droplets with unequal volumes ($V_1 \neq V_2$) or outer surface tensions ($\gamma_{\rm{f},1} \neq \gamma_{\rm{f},2}$), Eq.~\eqref{eq:contact_area} does not describe the size of the contact area. To derive the equilibrium shape and contact size of asymmetric droplets, we compute the minimum of the surface energy
\begin{align}
    E = \gamma_{\rm{c}} A_{\rm{c}} + \gamma_{\rm{f},1} A_{\rm{f},1} +  \gamma_{\rm{f},2} A_{\rm{f},2}
    \label{eq:energy_asymmetric_doublet}
\end{align}
under constant volume constraint. We follow the approach and use the parameterization introduced in \cite{maitre2016asymmetric}, which is shown in Fig.~\ref{fig:asymmetry_doublet_parameterization}(a). The droplet volumes can be expressed in terms of three spherical cap volumes $v_i$ with $i\in\{1,2,\rm{c}\}$ [Fig.~\ref{fig:asymmetry_doublet_parameterization}(b)] such that
\begin{align}
    V_1 = v_1 + v_c \\
    V_2 = v_2 - v_c.
\end{align}
Given the radii of curvature $R_i$ and the radius $r$ as shown in Fig.~\ref{fig:asymmetry_doublet_parameterization}(a), we can define the length scales
\begin{align}
    a_i = \sqrt{R_i^2 - r^2}
\end{align}
and surfaces
\begin{align}
    H_i(a_i, r) = \frac{1}{2}\left( a_i^2 + r^2 + a_i \sqrt{a_i^2 + r^2} \right),
\end{align}
which allows to express the spherical cap volumes as
\begin{align}
    v_i(a_i, r) = \frac{\pi}{3} \left(a_i + \sqrt{a_i^2 + r^2}\right)^2 \left(2 \sqrt{a_i^2 + r^2} - a_i \right)
\end{align}
and the different droplet surfaces as 
\begin{align}
    A_i(a_i, r) = 4 \pi H_i(a_i, r).
\end{align}
Using these definitions and expressing the outer surface tensions as $\gamma_{\rm{f},1} = \bar{\gamma}_{\rm{f}} + \delta \gamma_{\rm{f}}, \gamma_{\rm{f},2} = \bar{\gamma}_{\rm{f}} - \delta \gamma_{\rm{f}}$, we can rewrite Eq.~\eqref{eq:energy_asymmetric_doublet} as
\begin{align}
    \frac{E}{4 \pi \bar{\gamma}_{\rm{f}}} &= 
    \left(1 + \frac{\delta \gamma_{\rm{f}}}{\bar{\gamma}_{\rm{f}}} \right) H_1(a_1,r)
    +  \left(1 - \frac{\delta \gamma_{\rm{f}}}{\bar{\gamma}_{\rm{f}}} \right) H_2(a_2,r) \nonumber \\
     &+  2 \left(\frac{ \gamma_{\rm{c}} }{2 \bar{\gamma}_{\rm{f}}} \right) H_{\rm{c}}(a_{\rm{c}},r)
     \label{eq:energy_asymmetric_doublet_2}
\end{align}
with $\bar{\gamma}_{\rm{f}} = (\gamma_{\rm{f},1} + \gamma_{\rm{f},2})/2$. The minima in terms of the four parameters ($a_1,a_2,a_{\rm{c}},r$) under constant volume constraints $V_1 = \bar{V} - \delta V, V_2 = \bar{V} + \delta V$ were computed numerically, allowing to derive the size of the contact area $A_{\rm{c}}=4\pi H_c(a_c,r)$ (Appendix~\ref{sec:numerical_methods}).

\section{Numerical methods}

\begin{table*}[ht!]
    \centering
    \begin{tabular}{|p{70 mm}|p{20 mm}|p{60mm}|}
        \hline
         Physical quantity & Symbol & Values\\
         \hline
         Base line interfacial tension relative to outer surface tension in a single adherent droplet & $(\gamma_0-\gamma_{\rm{m}})/2\gamma_{\rm{f}}$ & Fig.~\ref{fig:3}(a,c): 0.95 \hfill\newline \\
         \hline
         Base line interfacial tension relative to outer surface tension & $\gamma_0/2\gamma_{\rm{f}}$ & 
         Fig.~\ref{fig:4}(a): 0.9 \hfill\newline 
         Fig.~\ref{fig:4}(c): 0.7 \hfill\newline
          Fig.~\ref{fig:5}-\ref{fig:7},\ref{fig:symmetry_breaking},\ref{fig:SteadyStateContactArea},\ref{fig:SPDetails}: 0.98 \hfill\newline
         \\
         \hline
         Reference susceptibility & $\chi_0$ & 
         Fig.~\ref{fig:3}: $\chi_0^{\rm{cusp}} = 2.8612$ \hfill \newline
         Fig.~\ref{fig:4}: $\chi_0^{\rm{PF}} = 16.629$ \hfill \newline
         Fig.~\ref{fig:5}-\ref{fig:7},\ref{fig:SteadyStateContactArea}: $\chi_0 = 40.604$
         \hfill \newline
         \\
         \hline
         Adaptive adhesion coefficient relative to outer surface tension & $\gamma_{\rm{A}}/2\gamma_{\rm{f}}$ & 
         Fig.~\ref{fig:3}(a, inlet): 0.9 \hfill \newline
         Fig.~\ref{fig:4}(a, inlet): 0.8 \hfill \newline
         Fig.~\ref{fig:5}(b): \{square, triangle: 0.15, quarterfoil:0.21, star:0.23, cross:0.2352, pentagon:0.5\} \hfill \newline
         
         Fig.~\ref{fig:5}(d): 0.8 \hfill \newline 
          Fig.~\ref{fig:5}(e,f): \{0.4, 0.6, 0.8\} 
          Fig.~\ref{fig:7}(b): 0.637 \hfill \newline
          Fig.~\ref{fig:7}(c): \{0.245, 0.637\} \hfill \newline\\
         \hline
         Relative signal susceptibility & $\chi/\chi_0$ &
         Fig.~\ref{fig:4}(a, inset): 2.8 \hfill \newline
         Fig.~\ref{fig:5}(b): \{square:0.1, quarterfoil:0.61, star:0.604, cross:0.6021, pentagon:0.6, triangle:0.95\} \hfill \newline
         Fig.~\ref{fig:5}(c): \{0.62,0.665,0.68\} \hfill \newline
         Fig.~\ref{fig:5}(d): \{0.4704, 0.7388\} \hfill \newline
         Fig.~\ref{fig:7}(c): \{0.3,0.465,0.4885 0.5,0.5221 0.6, 0.74,0.9 \}
         \\
         \hline
        Volume asymmetry & $\delta V/\bar{V}$ & Fig.~\ref{fig:6}: \{0.25, 0.5\} \\
         \hline
         Tension asymmetry & $\delta \gamma_{\rm{f}}/ \bar{\gamma}_{\rm{f}}$ & Fig.~\ref{fig:6}: \{0.25, 0.5\} \\
         \hline
         Hill coefficient & $h$ & 
         Fig.~\ref{fig:3}(a,b): 2 \hfill \newline
         Fig.~\ref{fig:4}(a-c): 2 \hfill \newline
         Fig.~\ref{fig:5}-\ref{fig:7}: 4 \hfill \newline
         \\
         \hline
    \end{tabular}
    \caption{Parameter values.}
    \label{tab:parameter_values_figures}
\end{table*}

\label{sec:numerical_methods}

\subsection{Bifurcation analysis}
The state and bifurcation diagrams presented in Fig.~\ref{fig:5}(a),(c),
were computed via continuation with the MATLAB-based software package MatCont \cite{dhooge2008new} (MatCont7p4 and MATLAB R2021a, scripts with details and numerical settings available at https://git.embl.de/dullwebe/dullweber2024. In general, initial fixpoints to initialize the continuation were computed by integration over time using the Integrator Method ode45. 

The saddle-node homoclinic (HSN) and associated non-central homoclinic to saddle-node (NCH) bifurcations [Fig.~\ref{fig:7}(a)] could only be obtained using the GUI-based version of MatCont7p4. For this, continuation of limit cycles was performed for $\gamma_{A}/\gamma_0=0.95$ and decreasing continuation parameter $\chi$ until the period reached a value close to 100 indicating the presence of a homoclinic causing period divergence. For different values of $\gamma_{A}/\gamma_0=0.95$, this point consistently coincides with the saddle-node bifurcation line. From the limit cycle of largest period, continuation of a HSN was initialized with InitStepsize 0.01, MinStepsize 0.05, MaxStepsize 0.1, MaxNewtonIters 3, MaxCorrIters 10, MaxTestIters 10, VarTolerance 1e-6, FunTolerance 1e-6, TestTolerance 1e-5, Adapt 1, MaxNumPoints 2000, CheckClosed 50 and Jacobian Increment 1e-05. Continuation of the HSN allows to detect the NCH codimension 2 point.\\

Results of the continuation were confirmed using simulations and analysis in Mathematica 13.0 (notebook with a step-by-step explanation of the analysis available at https://git.embl.de/dullwebe/dullweber2024. Specifically, we tested the number and types of stable attractors in different parameter regimes with simulations using NDSolve and ParametricNDSolve with the equation simplification method \emph{Residuals}. Fixpoints shown in the phase plots Fig.~\ref{fig:5}(b) were computed numerically in Mathematica from the intersections of nullclines. The SHET [Fig.~\ref{fig:5}(b)], HSN and Hom [Fig.~\ref{fig:7}(b)] were computed from simulation trajectories at parameter values very close to the bifurcation point. The oscillation amplitude [Fig.~\ref{fig:5}(d),(e)] and period  [Fig.~\ref{fig:5}(a)] were computed from the extrema of simulated trajectories, and checked against the dominant Fourier components.\\

\subsection{Timescale of symmetry-breaking}

The time of symmetry-breaking  $T_{\rm{sym}}$ [Fig.~\ref{fig:2}(b),blue curve] was computed as the simulation time (Mathematica) until 99\% of the steady state internal state difference $|u_1-u_2|$ is reached, starting from initial conditions $(u_1,u_2)=(0.01,0.02)$. The saddle  and its eigenvalues [Fig.~\ref{fig:2}(b),red crosses] were found numerically in Mathematica from the intersections of nullclines. Eigenvalues were normalized against the maximum saddle eigenvalue at $\chi/\chi_0^{\rm{PF}}=2, \gamma_{\rm{A}}/\gamma_0 = 1$.

\subsection{Treatment of asymmetric droplet shapes}
To obtain estimates of the equilibrium shapes of asymmetric droplets, we numerically computed the minimum of Eq.~\eqref{eq:energy_asymmetric_doublet_2} in terms of the four parameters ($a_1,a_2,a_{\rm{c}},r$) and under the constant volume constraints $V_1 = \bar{V} - \delta V, V_2 = \bar{V} + \delta V$ in Mathematica~\cite{Mathematica}(https://git.embl.de/dullwebe/dullweber2024). We computed the contact area $A_{\rm{c}}=4\pi H_c$ for values of $\gamma_{\rm{c}}/2\bar{\gamma}_{\rm{f}}$ evenly spaced on the interval $[0,1]$. From these results, we fit the contact area as a function of the tension ratio [Fig.~\ref{fig:asymmetry_doublet_parameterization}(c)], because our implementation of the numerical continuation method to obtain bifurcation lines requires an explicit expression that relates the contact area to the interfacial tensions. For unequal volumes ($V_1 \neq V_2$), but identical outer surface tensions, we used a 5th order polynomial to fit a function $A_{\rm{c}} = A_{\rm{c}}(\gamma_{\rm{c}}/2\bar{\gamma}_{\rm{f}})$ on the interval $[0,1]$ using Mathematica's function \emph{Fit} with the default \emph{LevenbergMarquardt} method [Fig.~\eqref{fig:asymmetry_doublet_parameterization}(c)].
For droplets with asymmetric outer tension, but equal volumes, the droplet with higher outer tension is completely internalized if $\gamma_{\rm{c}}/2\bar{\gamma}_{\rm{f}} \leq \delta \gamma_{\rm{f}}$ \cite{maitre2016asymmetric}, thus, we used a piecewise function to fit the contact area with $A_{\rm{c}} = 2^{4/3} A_0$ on the interval $[0,\delta \gamma_{\rm{f}}]$. The interval $[\delta \gamma_{\rm{f}},1]$ was fitted with a combination of a rational function of the form $a + b/(\gamma_{\rm{c}}/2\gamma_{\rm{f}} - c)^d$ close to the threshold of internalization with fit parameters $a-d$ and a 5th order polynomial [Fig.~\ref{fig:asymmetry_doublet_parameterization}(c)]. Fits of the contact area were then used for continuation in MatCont and simulations in Mathematica to derive the state diagrams shown in Fig.~\ref{fig:6}. 

All codes are available at https://git.embl.de/dullwebe/dullweber2024.

\section{Literature values for reaction and diffusion rates}

Khait et al. 2016 obtained quantitative estimates for many parameters governing the dynamics of Notch receptors and ligands (Tab.~\ref{tab:parameter_values}, \cite{khait2016quantitative}). The authors measured the 2D diffusion constant $D_{\mathrm{m}}$ and the endocytosis rate $k_{\rm{off}}$ in different cell lines and estimated the reaction rates from previously reported measurements of the binding kinetics for soluble molecules in 3D. The rate of receptor and ligand transport from the bulk to the surface $k_{\rm{on}}^{\rm{R}} c_{\rm{R}}^0$ and $k_{\rm{on}}^{\rm{L}} c_{\rm{L}}^0$ were estimated by assuming that the steady state surface densities are $m_{\rm{R}}^0=\SI{100}{\per\micro\metre\squared}$ and $m_{\rm{L}}^0=\SI{10}{\per\micro\metre\squared}$ (i.e. there is an excess of receptors) when a cell is not in contact with another cell or ligand-coated substrate. In that case $m_{\rm{R}}=m_{\rm{R}}^0=k_{\rm{on}}^{\rm{R}} c_{\rm{R}}^0/k_{\rm{off}}^{\rm{R}}$ and $m_{\rm{L}}=m_{\rm{L}}^0=k_{\rm{on}}^{\rm{L}} c_{\rm{L}}^0/k_{\rm{off}}^{\rm{L}}$.
$A_{\rm{c}}$ gives the range of contact areas for two spherical cells with radius $\SI{5}{\micro\metre}$ that keep a normalized contact area $A_{\rm{c}}/A_0 \in [0,1]$ as assumed in this work.

\begin{table}[h!]
    \centering
    \begin{tabular}{|c|c|c|}
    \hline
        Parameter&  Symbol& Value \\
        \hline 
        Endocytosis&  $k_{\rm{off}}$ & \SI{0.02}{\per\second}\\
        Cleavage &  $k_{\rm{s}}$ & \SI{0.34}{\per\second} \\
        Binding &  $k_+$ & \SI{0.167}{\micro\metre\squared\per\second}\\
        Unbinding &  $k_-$& \SI{0.034}{\per\second}\\
        Diffusion coefficients &   $D_{m_{\rm{R}}},  D_{m_{\rm{L}}}$ & \SIrange{0.02}{0.08}{\micro\metre\squared\per\second}\\
        Diffusion coefficient &  $D_{m_\mathrm{RL}}$ & $(D_{m_\mathrm{R}} + D_{m_\mathrm{L}})/4$\\
        Exocytosis receptors &  $k_{\rm{on}}^{\rm{R}} c_{\rm{R}}^0$ & \SI{2}{\per\micro\metre\squared\per\second}\\
        Exocytosis ligands &  $k_{\rm{on}}^{\rm{L}} c_{\rm{L}}^0$ & \SI{0.2}{\per\micro\metre\squared\per\second}\\
        Contact area &  $A_{\rm{c}}$ & \SIrange{0}{125}{\micro\metre\squared}\\
        \hline
    \end{tabular}
    \caption{Typical parameter values for reaction and diffusion rates of receptor and ligand molecules as reported in \cite{khait2016quantitative} and estimate of common cellular length scales}
    \label{tab:parameter_values}
\end{table}

\begin{figure*}[ht!]
    \centering
    \includegraphics[width=17.8cm]{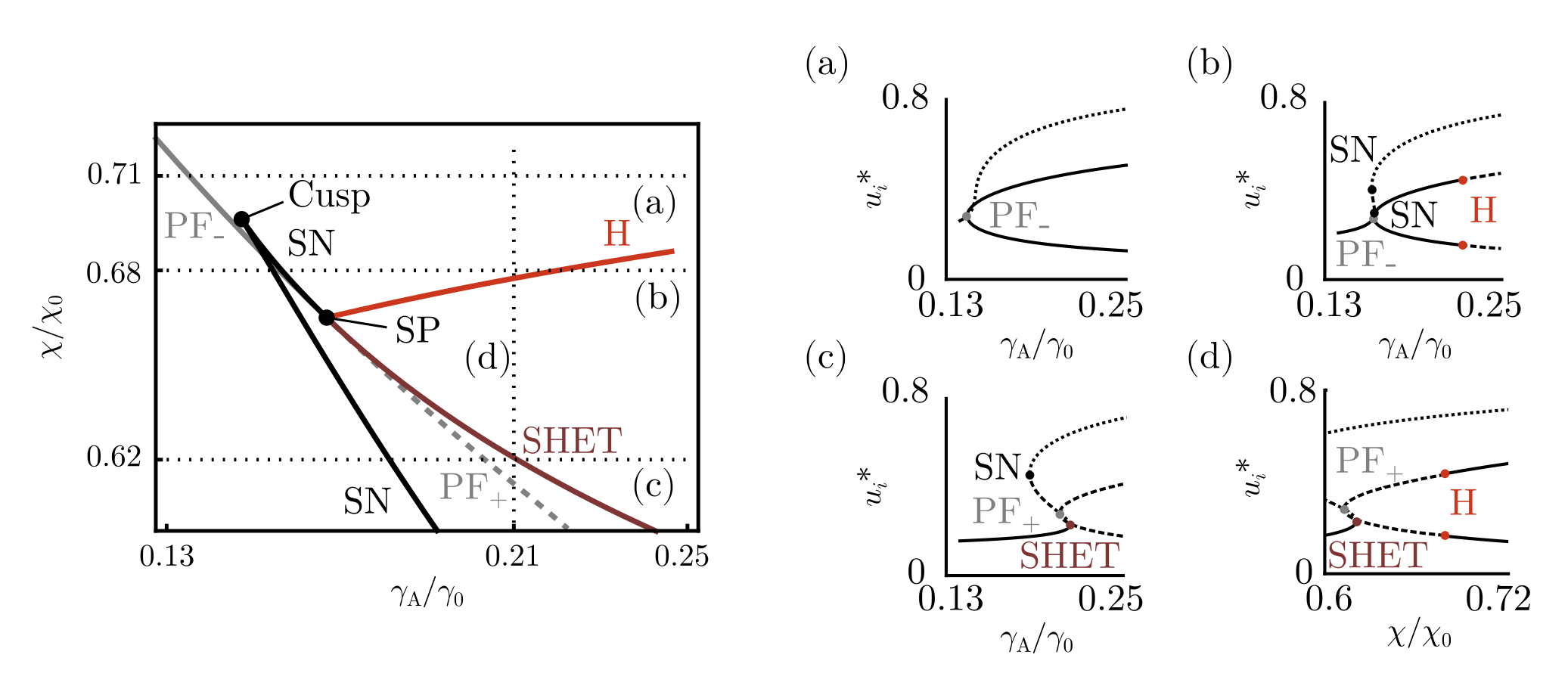}
    \caption{Bifurcation analysis close to the saddle-node pitchfork, 
    Left: Enlarged view of the state diagram of the doublet in terms of normalized feedback control parameters shown in Fig.~\ref{fig:5}(a) close to the saddle-node pitchfork (SP) codimension-2 bifurcation point.  (a) - (d) show stable (solid line) and unstable (dashed line) fixpoints and saddles (dotted line) computed for variation of one feedback parameter as indicated by dotted lines in the state diagram on the left.  As the pitchfork interacts with one of the saddle-nodes (compare (b) and (c)), it changes from supercritical (PF$_-$) to subcritical (PF$_+$) and the saddle (SN) becomes a Saddle-Heteroclinic (SHET). In the parameter regime between the H and SHET bifurcation lines, the system has no stable fixpoints, but stable limit cycles. H: Hopf bifurcation. Diagrams were computed in MatCont (Appendix~\ref{sec:numerical_methods}). This result is also presented in the supplementary material of the companion letter~\cite[Fig.~10]{PRLJoint}}
    \label{fig:SPDetails}
\end{figure*}

\section{Experimental methods (Fig.~\ref{fig:1}(a))}
\label{sec:experimental_methods}

\subsection{Cell culturing}
NIH/3T3 fibroblasts (ATCC CRL-1658, strain: NIH/Swiss) were cultured in \SI{10}{\centi\metre} plastic petri dishes at 37$^{\circ}$C and 5\% CO2 in phenol-red free DMEM medium containing 10\%FCS (in the following just called ”medium”). Cells were passaged every 2-3 days at $\sim~80-90\%$ confluency by washing with dPBS followed by \SIrange{2}{3}{\minute} incubation with \SI{1}{\milli\litre} Trypsin at 37$^{\circ}$C. Detached cells were resuspended in \SIrange{4}{5}{\milli\litre} medium and added either to a new \SI{10}{\centi\metre} petri dish with \SI{10}{\milli\litre} medium for culturing or a 6-well plate with \SI{2}{\milli\litre} medium per well for experiments, yielding a final confluency of 10-20\%.

\subsection{Imaging cells on micropatterns (Fig.~\ref{fig:1}(a))}
3T3 fibroblasts were seeded on Fibronectin-coated micropatterns of \SI{10}{\micro\metre} diameter as described in \cite{lembo2023distance}. Seeded cells were washed with dPBS and then phenol-red free DMEM medium with \SI{10}{\micro\gram\per\milli\litre} Hoechst33342 (1:1000 from \SI{10}{\milli\gram\per\milli\litre} in DMSO) and \SI{1}{\micro\mole} SiR-actin added to the dish. Cells were imaged $\geq\SI{30}{\minute}$ after adding the live dyes using a SP5 Confocal Microscope (Leica) with a 100x oil-immersion objective (1.4 NA).


\bibliography{references.bib}

\end{document}